\documentclass[10pt,aps,prd,onecolumn,superscriptaddress,amsmath,natbib, nofootinbib]{revtex4-2}

\DeclareUnicodeCharacter{2212}{-}

\RequirePackage{ifpdf}
\usepackage{amsmath}
\usepackage{mathtools} 

\usepackage{pstricks}
\usepackage{slashed}

\usepackage{hyperref}

\usepackage{color}
\usepackage{graphicx}
\usepackage{dcolumn}
\usepackage{bm}

\usepackage[mathlines]{lineno}

\usepackage{etoolbox} 
\usepackage{fixmath}

\usepackage[capitalise, nameinlink]{cleveref}
\usepackage{booktabs}
\usepackage{natbib}
\usepackage{cleveref}

\usepackage{notoccite} 

\usepackage{caption}
\usepackage{subcaption}
\usepackage{amsfonts}
\usepackage{multirow}

\usepackage{tabularx}
\usepackage{float}
\newcommand{\mk}[1]{{\color{black} #1 }}

\begin{document}

\preprint{}
\title{Next to SV resummed prediction for pseudoscalar Higgs boson production at NNLO$+\overline{\text{NNLL}}$}

\author{Arunima Bhattacharya}
\email{arunima.bhattacharya@saha.ac.in}
\affiliation{Saha Institute of Nuclear Physics, 1/AF Saltlake, Kolkata 700064, India}
\affiliation{Homi Bhabha National Institute, Training School Complex, Anushaktinagar, Mumbai 400094, India}
\author{M. C. Kumar}
\email{mckumar@iitg.ac.in}
\affiliation{Department of Physics, Indian Institute of Technology Guwahati, Guwahati-781039, Assam, India}
\author{Prakash Mathews}
\email{prakash.mathews@saha.ac.in}
\affiliation{Saha Institute of Nuclear Physics, 1/AF Saltlake, Kolkata 700064, India}
\affiliation{Homi Bhabha National Institute, Training School Complex, Anushaktinagar, Mumbai 400094, India}
\author{V.~Ravindran}
\email{ravindra@imsc.res.in}
\affiliation{The Institute of Mathematical Sciences, Taramani, Chennai-600113, India}
\affiliation{Homi Bhabha National Institute, Training School Complex, Anushaktinagar, Mumbai 400094, India}


\begin{abstract}
\noindent We present first results on the resummation of Next-to-Soft Virtual (NSV) logarithms for the threshold production of pseudoscalar Higgs boson through gluon fusion at the LHC. 
These results are presented after resumming the NSV logarithms of
the kind ${\log}^{i}(1-z)$ to $\overline{\text{NNLL}}$ accuracy and matching them systematically to the fixed order NNLO cross-sections.
These results are obtained using collinear factorization, renormalization group invariance and recent developments in the NSV resummation techniques.
The phenomenological implications of these NSV resummed results for 13 TeV LHC are studied and it is observed that these NSV logarithms are quite large. 
We also evaluate theory uncertainties and find that the 
renormalization scale uncertainties get reduced further with the inclusion of NSV corrections at various orders in QCD.
We further study the impact of QCD corrections on mixed scalar-pseudoscalar states for different values of the mixing angle $\alpha$.
\end{abstract}



\maketitle

\newcommand{\dis}[1]{\mathbold{#1}}
\newcommand{\overbar}[1]{mkern-1.5mu\overline{\mkern-1.5mu#1\mkern-1.5mu}\mkern
1.5mu}

\section{Introduction}
\label{sec:Introduction}

The ATLAS \cite{Aad:2012tfa} and CMS \cite{Chatrchyan:2012ufa} collaborations of the Large Hadron Collider (LHC) have been 
successful in discovering the Higgs boson of the standard model (SM) and this has put the SM on a very strong footing. As a 
result, a lot of work has been going on to investigate the properties and interactions of this discovered Higgs boson with the other
SM particles \cite{Higgs:1964ia,PhysRevLett.13.508,PhysRev.145.1156,PhysRevLett.13.321,PhysRevLett.13.585,ATLAS:2013sla,
ATLAS:2019nkf,2013,2015}. 
Despite this phenomenal success, it is widely known that the SM fails to explain certain natural phenomena such as the baryon 
asymmetry in the universe, existence of dark matter, tiny non-zero mass of neutrinos, etc. In order to explain these 
phenomena, one has to go beyond the realm of the SM. Supersymmetric theories provide one such solution to the above 
mentioned problems. The minimal supersymmetric extension of the SM (MSSM) is one of the simplest forms of the supersymmetric 
theories.  It has five Higgs bosons, out of which two are neutral scalars (h,H), one is a pseudoscalar (A) and the remaining 
two are charged scalars (H$^\pm$). The pseudoscalar Higgs boson which is CP odd could be as light as the discovered Higgs boson.
Hence, a dedicated effort has been going on to determine the CP property of the discovered Higgs boson, and to identify it 
with that of the SM, although there are already indications that it is a scalar with even parity \cite{2013,2015,
CMS:2012vby,ATLAS:2015bcp,ATLAS:2015zhl,CMS:2017len,ATLAS:2017azn}. 

\mk{Dedicated experimental searches for the pseudoscalar Higgs boson by both, the CMS and ATLAS collaborations,
have been carried out from the LHC data for $8$ TeV as well as $13$ TeV proton-proton collisions
\cite{CMS:2014ccx,ATLAS:2014vhc,ATLAS:2017nxi,Touquet:2019kho,Tsuno:2020jpd,ATLAS:2016ivh,CMS:2015grx,ATLAS:2017snw,CMS:2019pzc,ATLAS:2017eiz}. 
The searches for heavy scalar resonances decaying into a pair of $\tau$ leptons from the LHC 8 TeV data by CMS 
\cite{CMS:2014ccx} and ATLAS \cite{ATLAS:2014vhc} have excluded the values of $\tan \beta$ higher than $6.3~(57.6)$ for 
$m_A=100~(1000)$ GeV at 95\% confidence level. The experimental searches in the decay channel $A \to \tau \tau$ \cite{ATLAS:2016ivh}
have kept more stringent limits on the pseudoscalar parameter space than those obtained from the $A \to b \bar{b}$ 
channel \cite{CMS:2015grx}. The searches for pseudoscalar resonances in the top pair production by the ATLAS collaboration 
using data from the $8$ TeV LHC have put upper limits on $\tan \beta$ of the order of unity for $m_A > 500$ GeV 
\cite{ATLAS:2017snw}. Similar experimental searches by the CMS collaboration using data from $13$ TeV LHC have 
probed the pseudoscalar masses from $400$ GeV to $750$ GeV resulting in the exclusion of $\tan \beta$ values below $1.0$ to 
$1.5$ depending on the value of $m_A$ at 95\% confidence level \cite{CMS:2019pzc}.
The recent experimental searches for heavy resonances in the di-tau channel from the $13$ TeV LHC data by the ATLAS 
collaboration have excluded $\tan \beta > 1.0~(48)$ for $m_A = 250$ GeV ($1.5$ TeV) at 95\% confidence level \cite{ATLAS:2017eiz}.
In all these analyses, the higher-order corrections through the NNLO K-factor of about $2.0$ have been used. The large size of these corrections suggests that higher-order precision calculations through resummation can be useful in the experimental search for heavy scalar resonances. 
These studies, which have seen excesses over the background expectation in the Run 2 data of the LHC, indicate that extensive attempts are going on for discovering the BSM Higgs bosons with masses higher than the discovered Higgs boson mass of $125$ GeV. 
With such intense experimental works under process, developing the corresponding theoretical background has become necessary, and motivated us for this work.
}

\mk{There is also a possibility that the observed Higgs boson of $125$ GeV mass is an admixture of scalar and pseudoscalar 
states. If this is true, then such a mixed scalar can be produced in hadron collisions, through gluon fusion. Such a 
possibility has already been explored in \cite{Artoisenet:2013puc,Gao:2010qx,Maltoni:2013sma}. 
The identification of such a mixed scalar-pseudoscalar state is possible by studying various
kinematic distributions of the particles that this mixed state decays into \cite{Artoisenet:2013puc}.  
This requires the availability of fully differential distributions and such a study has been taken up at the NNLO level 
in \cite{Jaquier:2019bfs}. 
The existence of such a mixed state indicates possible new physics and hence, the CP violation in the Higgs sector. 
Moreover, it can also explain the origin of the CP violation in the SM and can address the problem of baryogenesis. 
All these imply the requirement of a detailed study for establishing the spin-parity properties of the discovered scalar boson of $125$ GeV mass. 
From the theory side, this necessitates precision studies for the relevant observable corresponding to both scalar and pseudoscalar production processes to the same order of precision.}

Higher order corrections in perturbative QCD (pQCD) provide a way to achieve the required precision. 
The pseudoscalar production cross-sections are available 
to NNLO accuracy in QCD \cite{Harlander_2002,Anastasiou_2003,Ravindran_2003}. 
The corrections are large and are of the order 
of 67$\%$ at NLO and get increased by an additional 15$\%$ at the NNLO level for the renormalization and factorization scales 
set to $\mu_R=\mu_F=m_A/2$ for the pseudoscalar mass $m_A = 200$ GeV. The large size of these corrections imply that in order to achieve 
precise theoretical results, corrections of even higher orders are necessary. 

The pseudoscalar Higgs production through gluon fusion channel via quark loop and the LO results for finite quark mass (exact) dependence are already available
\cite{Dawson:1990zj,Djouadi:1991tka,Spira:1995rr,Ravindran:2003um}.
The calculation becomes simpler in the infinite quark mass limit (effective field theory) 
and this makes it easy to compute cross-sections at higher orders in perturbation theory. 
This effective field theory (EFT) approach in the case of scalar Higgs boson production 
\cite{Anastasiou_2002,Harlander_2002h,Ravindran_2003} became extremely successful as the difference between the exact and EFT results  at NNLO level were found to be within 1\% \cite{Harlander_20091,Harlander_20092,Pak_2010,2021_Harlander}. 
For pseudoscalar production at the hadron colliders, NNLO predictions in the EFT approach are also available \cite{Ravindran_2003,Harlander_2002,Anastasiou_2003}. 
 
In \cite{Anastasiou:2014vaa,Anastasiou:2016cez} the computation of complete N$^{3}$LO predictions for the scalar Higgs boson 
production through gluon fusion at the hadron colliders in the effective theory has been accomplished. 
These third order corrections increase the cross-section by about $3.1\%$ for the central scale choice of $m_H/2$
while the corresponding scale uncertainty has reduced to as small as below $2\%$.
In a recent study for neutral current Drell-Yan process, the complete N$^3$LO results have been calculated for the first time \cite{Duhr:2021vwj}.
The corresponding cross-section is found to be about $0.992$ times that of
the NNLO cross-section for the invariant mass region $Q=300$ GeV indicating a small negative correction from the third order
in this kinematic region. Although the scale uncertainty at N$^3$LO level is found to be very mild, it has been observed
that the scale uncertainty bands for N$^3$LO and NNLO level cross-sections do not overlap with each other.
The next step in the process is the computation of 
N$^{3}$LO cross-sections for the pseudoscalar production through gluon fusion. The first task in this direction is to obtain the 
threshold enhanced cross-section at N$^{3}$LO level and the same has already been computed in \cite{Ahmed:2015qda}. 
Further,
the approximate full N$^3$LO$_A$ results are also available for the pseudoscalar Higgs boson case \cite{2016}.

However, these fixed order (FO) QCD predictions have limited applicability because of the presence of large logarithmic contributions that arise in the threshold region of pseudoscalar production. 
At the threshold, the emission of soft gluons gives large logarithmic contributions to 
the cross-section when the partonic center of mass energy approaches the pseudoscalar mass m$_A$. If these large logarithms arising from the soft gluons can be resummed to all orders in perturbation theory, then the problem of spoiling the reliability of FO perturbative predictions can be solved. We denote the soft-plus-virtual (SV) resummed results by LL, NLL, etc. The next-to-next-to-leading logarithmic (NNLL) resummed results \cite{2003,2005,2006,Ravindran_2006_B,Ravindran_2006,2006PRD,2009,2009PRL,2016PRD} give a sizable contribution and reduce scale uncertainties.
These logarithms in the parton level cross-section $d\hat{\sigma}$ computed to order $\alpha_s^k$ will appear as
\begin{equation}
 {\alpha_{s}}^k\left[\dfrac{\ln^{i}\left(1-z\right)}{1-z}\right]_{+}\text{, where }i<2k-1.
\end{equation}
Here, the subscript $+$ denotes the plus function and $z=m_A^2/s$ with $z\rightarrow1$ limit represents the partonic threshold region.  
The precise contribution of these parton level logarithms to the hadron level cross-section depends on the corresponding parton fluxes in that region.  
For the case of the discovered Higgs mass region $(125$ GeV$)$, the associated gluon flux is large 
and hence, these threshold logarithms due to the soft and collinear gluons are also found to be significant.
This way of resumming a set of large logarithms and then matching them to the FO results can give robust theoretical predictions.

For the scalar Higgs boson production, the FO results are available at N$^3$LO level 
\cite{Anastasiou_2015,Anastasiou:2016cez} and the corresponding threshold resummed results have been computed to N$^3$LL accuracy \cite{Bonvini:2014joa,Bonvini_2016_TROLL}. 
The computation of the three-loop threshold corrections for pseudoscalar Higgs boson production in the threshold limit has been done in \cite{Ahmed_2015}. 
A knowledge of the form factors up to three-loop level is needed to calculate the threshold corrections to pseudoscalar production at N$^3$LO level. 

Further, the ultraviolet (UV) and infrared (IR) divergences give rise to most of the logarithms at higher orders in the intermediate stages of the computations.  
These logarithms depend on the renormalization and factorization scales and are present in the perturbative expansions. 
Such logarithms help to estimate the error in theory predictions resulting from 
the truncation of the perturbation theory to a finite order. 

There are various studies on the computation of SV results in QCD corresponding to a number of observables produced in hadron collisions. For SV results up to third order, see \cite{Moch:2005ky,Ravindran_2006,de_Florian_2012,Ahmed:2014cla,Catani_2014,
Ravindran_2006_B,Ahmed:2014cha,Kumar:2014uwa,PhysRevD.90.053006}. 
A series of works in this direction have been carried out for the resummation of the threshold logarithms following the path breaking works by Sterman \textit{et al.} \cite{STERMAN1987310} and Catani \textit{et al.} \cite{CATANI1989323}.  
See \cite{CATANI1996273,Moch:2005ba,
Bonvini:2012an,Bonvini:2014joa,Bonvini:2014tea,Bonvini_2016_TROLL} for Higgs production in gluon fusion, 
\cite{Bonvini_2016,H:2019dcl} for bottom quark annihilation and for Drell-Yan (DY) 
\cite{Moch:2005ba,bonvini2010threshold,bonvini2012resummation,h2020resummed,Catani_2014}. 
In $z$-space, one has to deal with convolutions of these distributions and so, one moves to the Mellin space approach which uses the conjugate variable $N$ for resummation. The distributions $\mathcal{D}_i(z)$ become functions of $\text{log}^{i+1}(N)$ 
where we suppress terms of $\mathcal{O}(1/N)$ in the threshold limit of $N\rightarrow{\infty}$. 

However, it has also been observed that at higher orders in QCD, threshold corrections alone can not replace the full FO results and in fact, it is found that the role of the next-to soft-plus-virtual (NSV) terms, namely $\text{log}^{i}(1-z),~i=0,1,\cdots$, are also important. These NSV contributions can, in principle, originate from parton channels other than the one that corresponds to the Born contribution.  
It is also important to see whether these NSV terms 
can be resummed systematically to all orders exactly like the leading SV terms. Several advancements in this direction 
have been made \cite{Laenen:2008ux,Laenen:2010uz,Bonocore:2014wua,Bonocore:2015esa,Bonocore:2016awd,DelDuca:2017twk,Bahjat-Abbas:2019fqa,Moch:2009hr,Beneke:2018gvs,Beneke:2019mua,Beneke:2019oqx}. In \cite{ajjath2020soft,Ajjath:2022kyb}, the theories of mass factorization, renormalization group invariance and Sudakov K-plus-G equation have been exploited to provide a result in $z$- and $N$-space to predict the NSV terms for DY and Higgs boson production to all orders in perturbation theory. 
In \cite{2021}, the NSV resummation has been achieved to Leading Logarithmic ($\overline{\text{LL}}$) accuracy for color singlet production processes in hadron collisions. 
It has been observed that the resummation of NSV logarithms in the diagonal channel
gives large contributions to the cross-sections, while those from the $qg$ channel are found to give negative contributions of about $3\%$ in the high mass region. It has also been noticed that the inclusion of NSV resummed results increases scale uncertainties obtained from the SV resummation.

In this article, we compute the NSV corrections for the pseudoscalar Higgs boson production process at N$^3$LO level based on 
the formalism developed in \cite{Ravindran_2006,Ravindran_2006_B,ajjath2020soft,Ajjath:2022kyb} and the recent results at this order for scalar 
production \cite{Anastasiou:2016cez}. We further study the phenomenological impact of resumming these NSV logarithms to 
$\overline{\text{NNLL}}$ accuracy after they are systematically matched to the FO NNLO ones.
We represent the resummed results for the production of pseudoscalar Higgs boson at leading logarithmic, next-to leading logarithmic and next-to-next-to leading logarithmic accuracy by $\overline{\text{LL}}$, $\overline{\text{NLL}}$ and $\overline{\text{NNLL}}$ when we take into
account both SV and NSV threshold logarithms together.
The Section \ref{sec:TF} is devoted to developing the theoretical basis for this work. Section \ref{sec:TC} deals with the threshold 
corrections to the cross-section - the several ingredients required and subsequently, a detailed explanation of these elements. 
In Section \ref{sec:NSVres}, we present the analytical results for resumming the NSV logarithms to N$^3$LL accuracy. In Section \ref{sec:resNSVformalism} we recollect the NSV resummation formalism developed recently in \cite{ajjath2020soft,Ajjath:2022kyb} and in Section \ref{sec:Numerical} we present the numerical results for the production of pseudoscalar Higgs boson for 13 TeV LHC. 
Then we conclude this article.

\section{Theoretical Framework}
\label{sec:TF}

Coupling of a pseudoscalar Higgs boson to gluons happens through a virtual heavy
top quark loop indirectly which can be integrated out in the infinite
top quark mass limit. The effective lagrangian \cite{Chetyrkin_1998}
that is used to describe the interaction of the pseudoscalar fields, $\Phi^{A}(x)$,
with gauge fields \textit{via} the pseudoscalar gluonic operator,
$O_{G}(x)$, and with light quark fields \textit{via} the pseudoscalar
fermionic operator, $O_{J}(x)$, are given by
\begin{eqnarray}
{\cal L}_{eff}^{A}=\Phi^{A}(x)\left[-\frac{1}{8}C_{G}O_{G}(x)-\frac{1}{2}C_{J}O_{J}(x)\right]\,.
\label{eq:EL}
\end{eqnarray}
The pseudoscalar gluonic and fermionic operators are defined as 
\begin{align}
O_{G}(x) & =G^{a\mu\nu}\tilde{G}_{\mu\nu}^{a}=\varepsilon_{\mu\nu\rho\sigma}G^{a\mu\nu}G^{a\rho\sigma},\quad G^{a\mu\nu}=\partial^{\mu}G^{a\nu}-\partial^{\nu}G^{a\mu}+gf^{abc}G_{b}^{\mu}G_{c}^{\nu},\\
O_{J}(x) & =\partial_{\mu}\left(\bar{\psi}\gamma^{\mu}\gamma^{5}\psi\right)\,,
\end{align}
where $G_{\mu\nu}^{a}$ is the gluonic field strength tensor and $\epsilon_{\mu\nu\rho\sigma}$
is the Levi-Civita tensor. The pseudoscalar fermionic operator $O_{J}(x)$
is the derivative of the singlet axial vector current. The light quark field and its conjugate are represented by $\psi$ and $\bar{\psi}$ respectively. As we integrate out the heavy top quark degrees of freedom, the Wilson coefficients, $C_{G}$ and $C_{J}$, show a dependence on the mass of the top quark $m_{t}$. As a result of the Adler-Bardeen theorem \cite{PhysRev.177.2426}, there is no QCD corrections to $C_{G}$ beyond one-loop. On the other hand, $C_{J}$ begins only at second-order in the strong coupling constant. These coefficients are expanded in powers of the renormalized
strong coupling constant as a series of $a_{s}\equiv g_{s}^{2}/16\pi^{2}=\alpha_{s}/4\pi$.
The Wilson coefficients are given by 
\begin{align}
C_{G}\left(a_{s}\right) & =-a_{s}2^{\tfrac{5}{4}}G_{F}^{\tfrac{1}{2}}\cot\beta\,,\label{WCG}\\
C_{J}\left(a_{s}\right) & =-\left[a_{s}C_{F}\left(\dfrac{3}{2}-3\ln\dfrac{\mu_{R}^{2}}{m_{t}^{2}}\right)+a_{s}^{2}C_{J}^{\left(2\right)}+\cdots\right]C_{G}\,,\label{WCJ}
\end{align}
where $G_{F}$ is the Fermi constant, $\cot\beta$ is the ratio of
the two Higgs doublets' vacuum expectation values in a generic two-Higgs
doublet model, $C_{F}$ is the quadratic Casimir in the fundamental
representation of QCD and $\mu_{R}$ is the renormalization scale
at which $a_{s}$ is renormalized.

The unrenormalized strong coupling constant $\hat{a}_{s}$ is related to the renormalized one $a_s$ by
\begin{equation}
\hat{a}_{s}S_{\varepsilon}=\left(\dfrac{\mu^{2}}{\mu_{R}^{2}}\right)^{\varepsilon/2}Z_{a_{s}}a_{s},
\label{eq:ashattoas}
\end{equation}
with
\[
S_{\varepsilon}=\exp[(\gamma_E-\ln4\pi)\varepsilon/2],
\]
where the renormalization constant, $Z_{a_{s}}$, up to ${\cal{O}}(a_s^3)$ is given by \cite{TARASOV1980429,Ahmed_2015,Ravindran_2006}
\begin{equation}
 Z_{a_{s}}=1+a_{s}\left[\frac{2}{\epsilon}\beta_{0}\right]+a_{s}^{2}\left[\frac{4}{\epsilon^{2}}\beta_{0}^{2}+\frac{1}{\epsilon}\beta_{1}\right]+a_{s}^{3}\left[\frac{8}{\epsilon^{3}}\beta_{0}^{3}+\frac{14}{3\epsilon^{2}}\beta_{0}\beta_{1}+\frac{2}{3\epsilon}\beta_{2}\right]\,.
\end{equation}
Here $\mu$ is the scale introduced to keep the strong coupling
constant dimensionless in $d = 4 + \varepsilon$ space-time dimensions.
The QCD $\beta$ functions $(\beta_{i})$ have the standard definitions \cite{TARASOV1980429}.

\section{Threshold Corrections}
\label{sec:TC}

The inclusive cross-section for a pseudoscalar Higgs boson production
at the hadron colliders can be computed using \cite{Ahmed:2015qda}
\begin{equation}
\sigma^{A}\left(\tau,m_{A}^{2}\right)=\sigma^{A,\left(0\right)}\left(\mu_{R}^{2}\right)\sum_{a,b=q,\bar{q,g}}\int_{\tau}^{1}dy\ \Phi_{ab}\left(y,\mu_{F}^{2}\right)\Delta_{ab}^{A}\left(\dfrac{\tau}{y},m_{A}^{2},\mu_{R}^{2},\mu_{F}^{2}\right),\label{eq:sigmaA}
\end{equation}
where $\sigma^{A,\left(0\right)}\left(\mu_{R}^{2}\right)$ is the
born cross-section at the parton level with finite top mass dependence and is given by
\begin{equation}
\sigma^{A,\left(0\right)}\left(\mu_{R}^{2}\right) = \frac{\pi\sqrt{2}G_F}{16} a_s^2 \cot^2{\beta} |\tau_A f(\tau_A)|^2.
\end{equation}
In the above equation, $\tau_A=4m_t^2/m_A^2$ and the function $f(\tau_A)$ is given by
\begin{equation}
f(\tau_A) = 
     \begin{cases}
       \arcsin^2 \frac{1}{\sqrt{\tau_A}}&\quad\tau_A\geq1\\
       -\frac{1}{4}\bigg(\ln \frac{1-\sqrt{1-\tau_A}}{1+\sqrt{1-\tau_A}}+i\pi\bigg)^2 &\quad\tau_A<1 \\
     \end{cases}
\end{equation}
The parton flux is given by
\begin{equation}
\Phi_{ab}\left(y,\mu_{F}^{2}\right)=\int_{y}^{1}\dfrac{dx}{x}f_{a}\left(x,\mu_{F}^{2}\right)f_{b}\left(\dfrac{y}{x},\mu_{F}^{2}\right),
\end{equation}
where $f_{a}$ and $f_{b}$ are the parton distribution functions
(PDFs) of the initial state partons $a$ and $b$, renormalized at
the factorization scale $\mu_{F}$. Here, $\Delta_{ab}^{A}\left(\tau/y,m_{A}^{2},\mu_{R}^{2},\mu_{F}^{2}\right)$ represents the parton level cross-section for the sub-process initiated by $a$ and $b$ partons. This is the final result obtained after performing the 
UV renormalization at scale $\mu_R$ and the mass factorization at scale $\mu_F$.

The chief aim of this article is to study how the contribution from soft gluons
affects the cross-section for pseudoscalar production at hadron colliders. We 
get the final result that is IR safe by adding the soft part of the 
cross-section to the UV renormalized virtual part. However, even this 
is not enough as mass factorization is needed to be done using appropriate counter 
terms. This combination is what is known as the SV cross-section 
and the remaining part is known as the hard part. While the resummed results provide 
reliable predictions that can be compared against the experimental data, it is important 
to find out the role of sub-leading terms, namely $\log^{i}\left(1-z\right)$,
$i=0,1,\cdots$. We call them NSV contributions. 
Thus, we write the partonic cross-section as
\begin{equation}
\Delta_{gg}^{A}\left(z,q^{2},\mu_{R}^{2},\mu_{F}^{2}\right)=\Delta_{gg}^{A,NSV}\left(z,q^{2},\mu_{R}^{2},\mu_{F}^{2}\right)+\Delta_{gg}^{A,hard}\left(z,q^{2},\mu_{R}^{2},\mu_{F}^{2}\right),
\label{eq:deltaA}
\end{equation}
with $z\equiv q^{2}/\hat{s}=\tau/\left(x_{1}x_{2}\right)$. 
Here, $\Delta_{gg}^{A,NSV}\left(z,q^{2},\mu_{R}^{2},\mu_{F}^{2}\right)$ includes the threshold SV and NSV contributions, such that $\Delta_{gg}^{A,SV}\left(z,q^{2},\mu_{R}^{2},\mu_{F}^{2}\right)$ contains distributions of type $\delta\left(1-z\right)$ and $\mathcal{D}_{i}$ where the latter is defined as
\begin{equation}
\mathcal{D}_{i}\equiv\left[\dfrac{\ln^{i}\left(1-z\right)}{1-z}\right]_{+}.\label{eq:Distribution}
\end{equation}
On the contrary, the hard part of the cross-section, 
$\Delta_{gg}^{A,hard}\left(z,q^{2},\mu_{R}^{2},\mu_{F}^{2}\right)$,
contains all regular terms in $z$. 

The NSV cross-section in $z$-space is computed in $d=4+\varepsilon$ dimensions using \cite{ajjath2020soft}
\begin{equation}
\Delta_{gg}^{A,NSV}\left(z,q^{2},\mu_{R}^{2},\mu_{F}^{2}\right)=\mathcal{C}\exp\left\{ \varPsi_{g}^{A}\left(z,q^{2},\mu_{R}^{2},\mu_{F}^{2},\varepsilon\right)\right\} \mid_{\varepsilon=0}\label{eq:deltaNSV}
\end{equation}
where $\varPsi_{g}^{A}\left(z,q^{2},\mu_{R}^{2},\mu_{F}^{2},\varepsilon\right)$
is a finite distribution and $\mathcal{C}$ is the convolution defined as
\begin{equation}
\mathcal{C}e^{f\left(z\right)}=\delta\left(1-z\right)+\dfrac{1}{1!}f\left(z\right)+\dfrac{1}{2!}f\left(z\right)\otimes f\left(z\right)+\cdots. \label{eq:Convolution}
\end{equation}
Here $\otimes$ represents convolution and $f\left(z\right)$
is a distribution of the kind $\delta\left(1-z\right)$ and $\mathcal{D}_{i}$.
The subscript $g$ signifies the gluon initiated production of the pseudoscalar
Higgs boson. An equivalent formalism can be done in the Mellin ($N$-moment) space, which 
replaces the distributions in $z$ by continuous functions of the variable $N$. In this space, the threshold limit of $z\rightarrow1$ changes to $N\rightarrow\infty$. The finite distribution
$\varPsi_{g}^{A}$ depends on the form factors $\mathcal{F}_{g}^{A}\left(\hat{a}_{s},Q^{2},\mu^{2},\varepsilon\right)$ with $Q^2=-q^2$, the overall operator UV renormalization constant 
$Z_{g}^{A}\left(\hat{a_{s}},\mu_{R}^{2},\mu^{2},\varepsilon\right)$, the soft collinear distribution $\varPhi_{g}\left(\hat{a}_{s},q^{2},\mu^{2},z,\varepsilon\right)$ and the mass 
factorization kernels $\varGamma_{gg}\left(\hat{a}_{s},\mu_{F}^{2},\mu^{2},z,\varepsilon\right)$. 
The $\varPsi_{g}^{A}\left(z,q^{2},\mu_{R}^{2},\mu_{F}^{2},\varepsilon\right)$ can be written in terms of these quantities in the following form \cite{Ahmed:2015qda}:
\begin{align}
\varPsi_{g}^{A}\left(z,q^{2},\mu_{R}^{2},\mu_{F}^{2},\varepsilon\right)= & \left(\ln\left[Z_{g}^{A}\left(\hat{a_{s}},\mu_{R}^{2},\mu^{2},\varepsilon\right)\right]^{2}+\ln\mid\mathcal{F}_{g}^{A}\left(\hat{a}_{s},Q^{2},\mu^{2},\varepsilon\right)\mid\right)\delta\left(1-z\right)\nonumber \\
 & +\ 2\varPhi^{A}_{g}\left(\hat{a}_{s},q^{2},\mu^{2},z,\varepsilon\right)-2\mathcal{C}\ln\varGamma_{gg}\left(\hat{a}_{s},\mu_{F}^{2},\mu^{2},z,\varepsilon\right).
\end{align}

In the next few sections, we will elaborate on how to get these ingredients
to compute the NSV cross-section for pseudoscalar production at N$^{3}$LO in gluon fusion.

\subsection{Operator Renormalization Constant}

After the strong coupling constant renormalization through $Z_{a_{s}}$,
the form factor $\mathcal{F}_{g}^{A}\left(\hat{a}_{s},Q^{2},\mu^{2},\varepsilon\right)$ still does not become completely UV finite. 
The additional renormalization required to remove the residual UV divergences is called the overall operator renormalization and is done using the constant
$Z_{g}^{A}$. 
This is determined by solving the underlying RG equation \cite{Ahmed:2015qda}:
\begin{equation}
\mu_{R}^{2}\dfrac{d}{d\mu_{R}^{2}}\ln Z_{g}^{A}\left(\hat{a_{s}},\mu_{R}^{2},\mu^{2},\varepsilon\right)=\sum_{i=1}^{\infty}a_{s}^{i}\gamma_{g,i}^{A},
\end{equation}
where the UV anomalous dimensions ($\gamma_{g,i}^{A}$) upto three-loop $(i = 3)$ are as follows:
\begin{align}
\gamma_{g,1}^{A} & =\beta_{1},\\
\gamma_{g,2}^{A} & =\beta_{2},\\
\gamma_{g,3}^{A} & =\beta_{3}.\label{eq:gammaAgi}
\end{align}
Using the above renormalization group (RG) equation and the solutions of $\gamma_{g,i}^{A}$'s,
we obtain the overall renormalization constant up to three-loop level:
\begin{align}
Z_{g}^{A}= & 1+a_{s}\left[\dfrac{22}{3\varepsilon}C_{A}-\dfrac{4}{3\varepsilon}n_{f}\right] \nonumber \\ 
 & +a_{s}^{2}\bigg[\dfrac{1}{\varepsilon^{2}}\left\{ \dfrac{484}{9}C_{A}^{2}-\dfrac{176}{9}C_{A}n_{f}+\dfrac{16}{9}n_{f}^{2}\right\} + \dfrac{1}{\varepsilon}\left\{ \dfrac{34}{3}C_{A}^{2}-\dfrac{10}{3}C_{A}n_{f}-2C_{F}n_{f}\right\} \bigg]\ \nonumber \\ 
 & + a_{s}^{3}\bigg[\dfrac{1}{\varepsilon^{3}}\left\{ \dfrac{10648}{27}C_{A}^{3}-\dfrac{1936}{9}C_{A}^{2}n_{f}+\dfrac{352}{9}C_{A}n_{f}^{2}-\dfrac{64}{27}n_{f}^{3}\right\} \nonumber \\
 & +\dfrac{1}{\varepsilon^{2}}\left\{ \dfrac{5236}{27}C_{A}^{3}-\dfrac{2492}{27}C_{A}^{2}n_{f}-\dfrac{308}{9}C_{A}C_{F}n_{f}+\dfrac{280}{27}C_{A}n_{f}^{2}+\dfrac{56}{9}C_{F}n_{f}^{2}\right\} \nonumber \\
 & +\dfrac{1}{\varepsilon}\left\{ \dfrac{2857}{81}C_{A}^{3}-\dfrac{1415}{81}C_{A}^{2}n_{f}-\dfrac{205}{27}C_{A}C_{F}n_{f}+\dfrac{2}{3}C_{F}^{2}n_{f}+\dfrac{79}{81}C_{A}n_{f}^{2}+\dfrac{22}{27}C_{F}n_{f}^{2}\right\} \bigg].
 \label{eq:ZAg}
\end{align}
As can be seen, $Z_{g}^{A}=Z_{GG}$ where $Z_{GG}$ is the renormalization constant for $O_G$ operators which have been discussed in detail in \cite{Ahmed_2015,Bhattacharya_2020}.

\subsection{The Form Factor}

The unrenormalized form factor, $\hat{\mathcal{F}}_{g}^{A,(n)}$, can be expanded in terms of $\hat{a}_s$ as below:
\begin{equation}
  \label{eq:FF3}
  \mathcal{F}_{g}^{A} \equiv \sum_{n=0}^{\infty} \left[ {\hat{a}}_{s}^{n}
  \left( \frac{Q^{2}}{\mu^{2}} \right)^{n\frac{\varepsilon}{2}}
  S_{\varepsilon}^{n}  {\hat{\mathcal{F}}}^{A,(n)}_{g}\right] \,.
\end{equation}
The unrenormalized results for the scale choice of $\mu_{R}^{2}=\mu_{F}^{2}=q^{2}$ are given up to two-loop in \cite{Ravindran:2004mb} and up to three-loop in \cite{Ahmed:2015qda}. These are required in the context of computing NSV cross-section as discussed below.
The fact that QCD amplitudes obey factorization property, gauge and RG invariances lead to the consequence that the form factor 
$\mathcal{F}_{g}^{A}\left(\hat{a}_{s},Q^{2},\mu^{2},\varepsilon\right)$ satisfies
the following $K+G$ type differential equation \cite{Sudakov:1954sw, Mueller:1979ih, Collins:1980ih, Sen:1981sd, Magnea:1990zb, Ahmed:2015qda}:
%
\begin{equation}
Q^{2}\cfrac{d}{dQ^{2}}\ln\mathcal{F}_{g}^{A}\left(\hat{a}_{s},Q^{2},\mu^{2},\varepsilon\right)=\dfrac{1}{2}\left[K_{g}^{A}\left(\hat{a}_{s},\dfrac{\mu_{R}^{2}}{\mu^{2}},\varepsilon\right)+G_{g}^{A}\left(\hat{a}_{s},\dfrac{Q^{2}}{\mu_{R}^{2}},\dfrac{\mu_{R}^{2}}{\mu^{2}},\varepsilon\right)\right].
\end{equation}
All the poles in $\varepsilon$ are contained in the $Q^{2}$ independent function $K_{g}^{A}$ and those which are finite as $\varepsilon\rightarrow0$ are contained
in $G_{g}^{A}$. The solution of the above $K+G$ equation is given in a desirable form as below \cite{Ravindran_2006_B}:
%
\begin{equation}
\ln\mathcal{F}_{g}^{A}\left(\hat{a}_{s},Q^{2},\mu^{2},\varepsilon\right)=\sum_{i=1}^{\infty}\hat{a}_{s}^{i}\left(\dfrac{Q^{2}}{\mu^{2}}\right)^{i\dfrac{\varepsilon}{2}}S_{\varepsilon}^{i}\hat{\mathcal{L}}_{g,i}^{A}\left(\varepsilon\right),\label{eq:lnFAg}
\end{equation}
with
%
\begin{align}
\hat{\mathcal{L}}_{g,1}^{A}\left(\varepsilon\right)= & \dfrac{1}{\varepsilon^{2}}\left\{ -2A_{g,1}^{A}\right\} +\dfrac{1}{\varepsilon}\left\{ G_{g,1}^{A}\left(\varepsilon\right)\right\} ,\\
\hat{\mathcal{L}}_{g,2}^{A}\left(\varepsilon\right)= & \dfrac{1}{\varepsilon^{3}}\left\{ \beta_{0}A_{g,1}^{A}\right\} +\dfrac{1}{\varepsilon^{2}}\left\{ -\dfrac{1}{2}A_{g,2}^{A}-\beta_{0}G_{g,1}^{A}\left(\varepsilon\right)\right\} +\dfrac{1}{\varepsilon}\left\{ \dfrac{1}{2}G_{g,2}^{A}\left(\varepsilon\right)\right\} ,\\
\hat{\mathcal{L}}_{g,3}^{A}\left(\varepsilon\right)= & \dfrac{1}{\varepsilon^{4}}\left\{ -\dfrac{8}{9}\beta_{0}^{2}A_{g,1}^{A}\right\} +\dfrac{1}{\varepsilon^{3}}\left\{ \dfrac{2}{9}\beta_{1}A_{g,1}^{A}+\dfrac{8}{9}\beta_{0}A_{g,2}^{A}+\dfrac{4}{3}\beta_{0}^{2}G_{g,1}^{A}\left(\varepsilon\right)\right\} \nonumber \\
 & +\dfrac{1}{\varepsilon^{2}}\left\{ -\dfrac{2}{9}A_{g,3}^{A}-\dfrac{1}{3}\beta_{1}G_{g,1}^{A}\left(\varepsilon\right)-\dfrac{4}{3}\beta_{0}G_{g,2}^{A}\left(\varepsilon\right)\right\} +\dfrac{1}{\varepsilon}\left\{ \dfrac{1}{3}G_{g,3}^{A}\left(\varepsilon\right)\right\} ,
\end{align}  
where $A_{g,i}^{A}$'s are the cusp anomalous dimensions and $G_{g,i}^{A}\left(\varepsilon\right)$ are the resummation functions. The resummation functions, $G_{g,i}^{A}\left(\varepsilon\right)$, further decompose into $g^{A,i}_{g,j}$'s, and collinear $\left(B_{g}^{A}\right)$, soft $\left(f_{g}^{A}\right)$ and UV $\left(\gamma_{g}^{A}\right)$ anomalous dimensions \cite{Moch_2005,Ahmed:2015qda,Ravindran_2006}. 
Here, the soft anomalous dimensions, $f_{g}^{A}$'s, were introduced for the first time in \cite{Ravindran:2004mb}. The article \cite{Ravindran:2004mb} has shown that 
$f_{g}^{A}$'s fulfill the maximally non-Abelian property up to two-loop level whose validity is reconfirmed in \cite{Moch_2005} at the three-loop level. 
All these $f_{g}$ can be found in \cite{Ravindran:2004mb, Moch_2005}, $A_{g,i}$ in~\cite{Moch:2004pa,Moch_2005,Vogt:2004mw,2021_Blumlein,Vogt:2000ci} and $B_{g,i}$ in \cite{Vogt:2004mw,2021_Blumlein,Moch_2005} up to three-loop level.

For computing the NSV cross-section, we need the $g^{A,i}_{g,j}$'s appearing in the $G_{g,i}^{A}\left(\varepsilon\right)$ resummation functions. For N$^3$LO calculations, 
$g^{A,1}_{g,3}$ is needed in addition to the quantities arising from the one- and two-
loops. The form factors for pseudoscalar production can be found in~\cite{Ravindran:2004mb} up to two-loop and is calculated by some of us in the article~\cite{Ahmed_2015}  up to three-loop. However, in this computation of the SV+NSV cross-section at N$^{3}$LO,
the form factor is needed in a definite form which is slightly different 
from the one presented in this recent article \cite{Ahmed_2015}. With a little bit of effort, the required form can be extracted from the above mentioned recent work. The $g_{g,i}^{A,k}$'s up to three-loop level are given below \cite{Ahmed:2015qda}:
\begin{align}
  \label{eq:g31}
  g^{A,1}_{g,1} &= {\dis{C_{A}}} \Bigg\{ 4 + \zeta_2 \Bigg\}\,,
                  \nonumber\\
  g^{A,2}_{g,1} &= {\dis{C_{A}}} \Bigg\{ - 6 - \frac{7}{3} \zeta_3 \Bigg\}\,,
                  \nonumber\\
  g^{A,3}_{g,1} &= {\dis{C_{A}}} \Bigg\{ 7 - \frac{1}{2} \zeta_2 + \frac{47}{80} \zeta_2^2 \Bigg\}\,,
                  \nonumber\\ \nonumber\\
  g^{A,1}_{g,2} &= {\dis{C_{A}^2}} \Bigg\{ \frac{11882}{81} + \frac{67}{3}
                  \zeta_2 - \frac{44}{3} \zeta_3 \Bigg\} 
                  + 
                  {\dis{C_{A} n_{f}}} \Bigg\{ - \frac{2534}{81} - \frac{10}{3} \zeta_2 -
                  \frac{40}{3} \zeta_3 \Bigg\} 
                  \nonumber\\
                & + {\dis{C_{F} n_{f}}} \Bigg\{ - \frac{160}{3} 
                  + 12 \ln \left(\frac{\mu_R^2}{m_t^2}\right) + 16 \zeta_3 \Bigg\}\,,
                  \nonumber\\
  g^{A,2}_{g,2} &= {\dis{C_{F} n_{f}}} \Bigg\{ \frac{2827}{18} - 18
                  \ln \left(\frac{\mu_R^2}{m_t^2}\right)  - \frac{19}{3} \zeta_2 - \frac{16}{3}
                  \zeta_2^2  - 
                  \frac{128}{3} \zeta_3 \Bigg\} 
                  + {\dis{C_{A} n_{f}}} \Bigg\{
                  \frac{21839}{243}  - \frac{17}{9} \zeta_2 
                  \nonumber\\
                &+
                  \frac{259}{60} \zeta_2^2  + 
                  \frac{766}{27} \zeta_3 \Bigg\} 
                  + {\dis{C_{A}^2}} \Bigg\{ -
                  \frac{223861}{486}  + \frac{80}{9} \zeta_2 +
                  \frac{671}{120} \zeta_2^2  + 
                  \frac{2111}{27} \zeta_3 + \frac{5}{3} \zeta_2
                  \zeta_3  - 39 \zeta_5 \Bigg\}\,,
                  \nonumber\\ \nonumber\\                  
  g^{A,1}_{g,3} &=  {\dis{n_{f} C_{J}^{(2)}}}  \Bigg\{ - 6 \Bigg\}  
                  + {\dis{C_{F}
                  n_{f}^2}} \Bigg\{ \frac{12395}{27}  -
                  \frac{136}{9} \zeta_2  - 
                  \frac{368}{45} \zeta_2^2 - \frac{1520}{9} \zeta_3  -
                  24  \ln \left(\frac{\mu_R^2}{m_t^2}\right) 
                  \Bigg\}  
                  \nonumber\\
                &+ 
                  {\dis{C_{F}^2 n_{f}}} \Bigg\{ \frac{457}{2} + 312 \zeta_3 -
                  480 \zeta_5 \Bigg\}  
                  + 
                  {\dis{C_{A}^2 n_{f}}} \Bigg\{ - \frac{12480497}{4374} -
                  \frac{2075}{243} \zeta_2  - \frac{128}{45} \zeta_2^2 
                  \nonumber\\                  
                & -\frac{12992}{81} \zeta_3 - \frac{88}{9} \zeta_2
                  \zeta_3 + \frac{272}{3} \zeta_5 \Bigg\}
                  + 
                  {\dis{C_{A}^3}} \Bigg\{ \frac{62867783}{8748} +
                  \frac{146677}{486} \zeta_2  - \frac{5744}{45}
                  \zeta_2^2  - 
                  \frac{12352}{315} \zeta_2^3 
                  \nonumber\\ 
                &- \frac{67766}{27}
                  \zeta_3 - \frac{1496}{9} \zeta_2 \zeta_3  - 
                  \frac{104}{3} \zeta_3^2 + \frac{3080}{3} \zeta_5 \Bigg\}
                  + 
                  {\dis{C_{A} n_{f}^2}} \Bigg\{ \frac{514997}{2187} -
                  \frac{8}{27} \zeta_2  + \frac{232}{45} \zeta_2^2 
                  \nonumber\\                  
                &+ 
                  \frac{7640}{81} \zeta_3 \Bigg\} 
                  + {\dis{C_{A} C_{F} n_{f}}} \Bigg\{
                  - \frac{1004195}{324}  + \frac{1031}{18} \zeta_2 + 
                  \frac{1568}{45} \zeta_2^2 + \frac{25784}{27} \zeta_3
                  + 40 \zeta_2 \zeta_3  + \frac{608}{3} \zeta_5 
                  \nonumber\\
                &+ 132 \ln \left(\frac{\mu_R^2}{m_t^2}\right) \Bigg\}\,.                
\end{align}

\subsection{Mass Factorization Kernel}

The partonic cross-section $\Delta_{gg}^{A,NSV}\left(z,q^{2},\mu_{R}^{2},\mu_{F}^{2}\right)$ is UV finite after performing the coupling constant and overall operator
renormalization using $Z_{a_{s}}$ and $Z_{g}^{A}$. 
But it still exhibits collinear divergences and thus, requires mass factorization to remove them. In this section, the issue of collinear divergences is dealt with and we describe a prescription to remove them. 
The collinear singularities that arise in the massless limit of partons are removed in the $\overline{MS}$ scheme using the mass factorization kernel $\varGamma\left(\hat{a}_{s},\mu^{2},\mu_{F}^{2},z,\varepsilon\right)$.
The kernel satisfies the following RG equation \cite{Ravindran_2006_B,Ahmed:2015qda}:
\begin{equation}
\mu_{F}^{2}\dfrac{d}{d\mu_{F}^{2}}\varGamma\left(z,\mu_{F}^{2},\varepsilon\right)=\dfrac{1}{2}P\left(z,\mu_{F}^{2}\right)\otimes\varGamma\left(z,\mu_{F}^{2},\varepsilon\right),
\end{equation}
where $P\left(z,\mu_{F}^{2}\right)$ are the Altarelli-Parisi (AP) splitting
functions (matrix valued). We can expand $P\left(z,\mu_{F}^{2}\right)$
and $\varGamma\left(z,\mu_{F}^{2},\varepsilon\right)$ in powers of
the strong coupling constant $a_{s}$ as follows:
\begin{equation}
P\left(z,\mu_{F}^{2}\right)=\sum_{i=1}^{\infty}a_{s}^{i}\left(\mu_{F}^{2}\right)P^{\left(i-1\right)}\left(z\right),
\end{equation}
and 
\begin{equation}
\varGamma\left(z,\mu_{F}^{2},\varepsilon\right)=\delta\left(1-z\right)+\sum_{i=1}^{\infty}\hat{a}_{s}^{i}\left(\dfrac{\mu_{F}^{2}}{\mu^{2}}\right)S_{\varepsilon}^{i}\varGamma^{\left(i\right)}\left(z,\varepsilon\right).
\end{equation}
We can solve the RGE for these mass factorization kernels. The solutions in the $\overline{MS}$ scheme contains only the poles in $\varepsilon$ and are given in \cite{Ravindran_2006_B}. The relevant values of the observables that are required for this computation are available in the articles~\cite{Moch:2004pa,Vogt:2004mw,2021_Blumlein}. Only the diagonal AP kernels contribute to our analysis. So, we expand the corresponding AP splitting functions around $z = 1$ and all those terms that do not contribute to NSV are dropped.

The AP splitting functions near $z = 1$ for the gluon fusion sub-process take the following form \cite{ajjath2020soft}:
\begin{align}
P_{gg,i}\left(z,a_{s}\left(\mu_{F}^{2}\right)\right)= & 2\bigg[B_{g,i}\left(a_{s}\left(\mu_{F}^{2}\right)\right)\delta\left(1-z\right)+A_{g,i}\mathcal{D}_{0}\left(z\right)\nonumber \\
 & +C_{g,i}\left(a_{s}\left(\mu_{F}^{2}\right)\right)\log\left(1-z\right)+D_{g,i}\left(a_{s}\left(\mu_{F}^{2}\right)\right)\bigg]+\mathcal{O}\left(1-z\right) ,
\end{align}
where $C_{g,i}$ and $D_{g,i}$ are constants that can be obtained from
the splitting functions $P_{gg,i}$. Just as the cusp and the collinear anomalous dimensions were expanded in powers of $a_s(\mu_F^2)$, the constants $C_g$ and $D_g$ can also be expanded similarly as below:
\begin{eqnarray}
C_g(a_s(\mu_F^2)) = \sum_{i=1}^\infty a_s^i(\mu_F^2) C_{g,i},
\quad \quad 
D_g(a_s(\mu_F^2)) = \sum_{i=1}^\infty a_s^i(\mu_F^2) D_{g,i} \,,
\end{eqnarray}
where $C_{g,i}$ and $D_{g,i}$ to third order are available in \cite{Moch:2004pa,Vogt:2004mw,2021_Blumlein}.

\subsection{Soft Collinear Distribution}
\label{subsec:SCD}

The resulting expression obtained after using the operator renormalization
constant and the mass factorization kernel is still not completely finite. It
contains some residual divergences which get cancelled against
the contribution arising from soft gluon emissions. This is why the
finiteness of $\Delta_{gg}^{A,NSV}\left(z,q^{2},\mu_{R}^{2},\mu_{F}^{2}\right)$
in the limit $\varepsilon\rightarrow0$ requires the soft-collinear
distribution $\varPhi_{g}\left(\hat{a}_{s},q^{2},\mu^{2},z,\varepsilon\right)$
which has a pole structure in $\varepsilon$ similar to that of the residual
divergences. 
The distribution $\varPhi_{g}\left(\hat{a}_{s},q^{2},\mu^{2},z,\varepsilon\right)$ 
satisfies the $K+G$ type differential equation \cite{ajjath2020soft}:
\begin{equation}
q^{2}\dfrac{d}{dq^{2}}\varPhi_{g}=\dfrac{1}{2}\left[\overline{K_{g}}\left(\hat{a}_{s},\dfrac{\mu_{R}^{2}}{\mu^{2}},\varepsilon,z\right)+\overline{G_{g}}\left(\hat{a}_{s},\dfrac{q^{2}}{\mu_{R}^{2}},\dfrac{\mu_{R}^{2}}{\mu^{2}},\varepsilon,z\right)\right],
\end{equation}
where $\overline{K_{g}}$ contains all the divergent terms and $\overline{G_{g}}$
contains all the finite functions of $\left(z,\varepsilon\right)$. 
A detailed study has been given in \cite{ajjath2020soft}.

We can rewrite $\varPhi_{g}\left(\hat{a}_{s},q^{2},\mu^{2},z,\varepsilon\right)$
in a convenient form which separates the SV terms from the NSV ones. 
Hence, we decompose it as 
\begin{equation}
\varPhi_{g}=\varPhi_{g}^{SV}+\varPhi^{NSV}_{g},
\end{equation}
such that $\varPhi^{SV}_{g}$ contains only the SV terms and the remaining $\varPhi^{NSV}_{g}$ contains the NSV terms in the limit $z\rightarrow1$. 
The form of $\varPhi^{SV}_{g}$ is given by
\begin{equation}
\varPhi^{SV}_{g}\left(\hat{a}_{s},q^{2},\mu^{2},z,\varepsilon\right)=\sum_{i=1}^{\infty}\hat{a}_{s}^{i}\left(\dfrac{q^{2}\left(1-z\right)^{2}}{\mu^{2}}\right)^{i\dfrac{\varepsilon}{2}}S_{\varepsilon}^{i}\left(\dfrac{i\varepsilon}{1-z}\right)\hat{\phi}_{g}^{SV,\left(i\right)}\left(\varepsilon\right),
\label{eq:phicapg}
\end{equation}
where the expressions for $\hat{\phi}_{g}^{SV,\left(i\right)}\left(\varepsilon\right)$'s are explicitly given in Appendix \ref{appendix:B}.

As verified by some of us (upto N$^3$LO) in \cite{Ahmed:2015qda}, due to the universality of the soft gluon contribution, 
$\varPhi_{g}^{SV}\left(\hat{a}_{s},q^{2},\mu^{2},z,\varepsilon\right)$ must be the same as
that of the Higgs boson production in gluon fusion:
\begin{align}
  \label{eq:PhiAgPhiHg}
  \varPhi^{A}_{g} &= \varPhi^{H}_{g} = \varPhi_{g},
                 \nonumber\\
  \text{i.e.}~~ {\overline {\cal G}}^{A,k}_{g,i} &=  {\overline {\cal G}}^{H,k}_{g,i} =  {\overline {\cal G}}^{k}_{g,i}\,. 
\end{align}
Here, ${\Phi}^{H}_{g}$ and $ {\overline {\cal G}}^{H,k}_{g,i}$ can be used for any gluon fusion process as these are independent of the operator insertion. 
The constants, ${\overline {\cal G}}_{g,1}^{H,1}, {\overline {\cal G}}_{g,1}^{H,2}, 
{\overline {\cal G}}_{g,2}^{H,1}$, were determined from the result of the explicit computations of soft gluon emissions to the Higgs boson production in \cite{Ravindran:2003um} and later, these corrections were further extended to all orders in the dimensional
regularization parameter $\varepsilon$ in \cite{de_Florian_2012},
using which $ {\overline {\cal G}}_{g,1}^{H,3}$ and $ {\overline {\cal G}}_{g,2}^{H,2}$ 
are extracted in \cite{Ahmed:2015qda}. 
\mk{A detailed description of these constants, the ${\overline {\cal G}}^{H,k}_{g,i}$'s or the 
${\overline {\cal G}}^{A,k}_{g,i}$'s, that are used in the evaluation of 
$\varPhi_{g}^{SV}\left(\hat{a}_{s},q^{2},\mu^{2},z,\varepsilon\right)$, are already given in \cite{Ahmed:2015qda}.  
Hence, we do not repeat them here but give the final results for $\hat{\phi}_{g}^{SV,\left(i\right)}\left(\varepsilon\right)$ in Appendix \ref{appendix:B} after applying the available relevant expressions.}
The third order constant $ {\overline {\cal G}}_{g,3}^{H,1}$ is computed from the result of SV cross-section for the production of the Higgs boson at N$^{3}$LO \cite{Anastasiou:2014vaa} which was presented in the article~\cite{Ahmed:2014cla}. 
%

Having discussed about the computation of the SV part of the cross-section, we now move on to calculate the corresponding NSV part. 
The form of $\varPhi_{g}^{NSV}$ is given by \cite{ajjath2020soft},
\begin{equation}
\varPhi_{g}^{NSV}\left(\hat{a}_{s},q^{2},\mu^{2},z,\varepsilon\right)=\sum_{i=1}^{\infty}\hat{a}_{s}^{i}\left(\dfrac{q^{2}\left(1-z\right)^{2}}{\mu^{2}}\right)^{i\dfrac{\varepsilon}{2}}S_{\varepsilon}^{i}\varphi_{g}^{NSV,\left(i\right)}\left(z,\varepsilon\right).
\end{equation}
The $\varphi_{g}^{NSV,\left(i\right)}\left(z,\varepsilon\right)$ coefficients can be expressed as a sum of singular and finite part in $\varepsilon$ given by,
\begin{equation}
\varphi_{g}^{NSV,\left(i\right)}\left(z,\varepsilon\right)=\varphi_{s,g}^{NSV,\left(i\right)}\left(z,\varepsilon\right)+\varphi_{f,g}^{NSV,\left(i\right)}\left(z,\varepsilon\right),
\label{eq:phiNSV}
\end{equation}
where explicit expressions for the singular coefficients $\varphi_{s,g}^{NSV,\left(i\right)}\left(z,\varepsilon\right)$ are given in Appendix \ref{appendix:C}.

The coefficients $\varphi_{f,g}^{NSV,\left(i\right)}\left(z,\varepsilon\right)$
are finite as $\varepsilon\rightarrow0$ and can be written in terms
of the finite coefficients $\mathcal{G}_{L,i}^{g}\left(z,\varepsilon\right)$
as \cite{ajjath2020soft},
\begin{align}
\varphi_{f,g}^{NSV,\left(1\right)}\left(z,\varepsilon\right)= & \dfrac{1}{\varepsilon}\mathcal{G}_{L,1}^{g}\left(z,\varepsilon\right),\\
\varphi_{f,g}^{NSV,\left(2\right)}\left(z,\varepsilon\right)= & \dfrac{1}{\varepsilon^{2}}\left\{ -\beta_{0}\mathcal{G}_{L,1}^{g}\left(z,\varepsilon\right)\right\} +\dfrac{1}{2\varepsilon}\mathcal{G}_{L,2}^{g}\left(z,\varepsilon\right),\\
\varphi_{f,g}^{NSV,\left(3\right)}\left(z,\varepsilon\right)= & \dfrac{1}{\varepsilon^{3}}\left\{ \dfrac{4}{3}\beta_{0}^{2}\mathcal{G}_{L,1}^{g}\left(z,\varepsilon\right)\right\} +\dfrac{1}{\varepsilon^{2}}\left\{ -\dfrac{1}{3}\beta_{1}\mathcal{G}_{L,1}^{g}\left(z,\varepsilon\right)-\dfrac{4}{3}\beta_{0}\mathcal{G}_{L,2}^{g}\left(z,\varepsilon\right)\right\} \nonumber\\
& +\dfrac{1}{3\varepsilon}\mathcal{G}_{L,3}^{g}\left(z,\varepsilon\right),
\end{align}
where 
\begin{align}
\mathcal{G}_{L,1}^{g}\left(z,\varepsilon\right)= & \sum_{j=1}^{\infty}\varepsilon^{j}\mathcal{G}_{L,1}^{g,\left(j\right)}\left(z\right),\\
\mathcal{G}_{L,2}^{g}\left(z,\varepsilon\right)= & -2\beta_{0}\mathcal{G}_{L,1}^{g,\left(1\right)}\left(z\right)+\sum_{j=1}^{\infty}\varepsilon^{j}\mathcal{G}_{L,2}^{g,\left(j\right)}\left(z\right),\\
\mathcal{G}_{L,3}^{g}\left(z,\varepsilon\right)= & -2\beta_{1}\mathcal{G}_{L,1}^{g,\left(1\right)}\left(z\right)-2\beta_{0}\left(\mathcal{G}_{L,2}^{g,\left(1\right)}\left(z\right)+2\beta_{0}\mathcal{G}_{L,1}^{g,\left(2\right)}\left(z\right)\right)+\sum_{j=1}^{\infty}\varepsilon^{j}\mathcal{G}_{L,3}^{g,\left(j\right)}\left(z\right).
\label{eq:curlyGg}
\end{align}
The coefficients $\mathcal{G}_{L,i}^{g,\left(j\right)}\left(z\right)$ given 
in the above equations are parameterized in terms of $\log^{k}\left(1-z\right),
k = 0,1,\cdots$ while all other terms that vanish as $z\rightarrow1$ are dropped
\begin{equation}
\mathcal{G}_{L,1}^{g,\left(j\right)}\left(z\right)=\sum_{k=0}^{i+j-1}\mathcal{G}_{L,i}^{g,\left(j,k\right)}\left(z\right)\log^{k}\left(1-z\right).
\end{equation}
The highest power of the log$\left(1-z\right)$ terms at every order depends
on the order of the perturbation i.e. the power of $a_{s}$ and also
the power of $\varepsilon$ at each order in $a_{s}$. 

The expansion coefficients $\varphi_{g,i}^{\left(k\right)}$ are related to $\mathcal{G}_{L,i}^{g,\left(j,k\right)}$ as below \cite{ajjath2020soft}:

\begin{align}
\label{eq:phigk1}  
\varphi_{g,1}^{\left(k\right)}= & \mathcal{G}_{L,1}^{g,\left(1,k\right)},\ \ \ \ \ \ k=0,1\\ 
\varphi_{g,2}^{\left(k\right)}= & \dfrac{1}{2}\mathcal{G}_{L,2}^{g,\left(1,k\right)}+\beta_{0}\mathcal{G}_{L,1}^{g,\left(2,k\right)},\ \ \ \ \ \ k=0,1,2\\ 
\varphi_{g,3}^{\left(k\right)}= & \dfrac{1}{3}\mathcal{G}_{L,3}^{g,\left(1,k\right)}+\dfrac{2}{3}\beta_{1}\mathcal{G}_{L,1}^{g,\left(2,k\right)}+\dfrac{2}{3}\beta_{0}\mathcal{G}_{L,2}^{g,\left(2,k\right)}+\dfrac{4}{3}\beta_{0}^{2}\mathcal{G}_{L,1}^{g,\left(3,k\right)},\ \ \ \ \ \ k=0,1,2,3.
\label{eq:phigk3}
\end{align}

\section{Next to SV results}
\label{sec:NSVres}

Using all the above available ingredients, we can calculate the NSV coefficient functions for the pseudoscalar Higgs boson production from gluon fusion in terms of the expansion coefficients, $\varphi_{g,i}^{\left(k\right)}$'s, defined in Section \ref{subsec:SCD}. 
The SV corrections to the production of pseudoscalar Higgs boson is available up to order $a_{s}^{3}$ while the corresponding NSV corrections are available up to order $a_{s}^{2}$. 
On the contrary, the NSV corrections to the production of scalar Higgs boson are available up to order $a_{s}^{3}$. 
As given in \cite{2016}, the similarity between pseudoscalar Higgs boson production and scalar Higgs boson production is exploited which leads to the conclusion that the pseudoscalar result can be approximated from the available scalar Higgs boson result using
\begin{align}
\Delta_{gg}^{A}\left(z,q^{2},\mu_{R}^{2},\mu_{F}^{2}\right)=& \dfrac{g_{0}\left(a_{s}\right)}{g_{0}^{H}\left(a_{s}\right)}\bigg[\Delta_{gg}^{H}\left(z,q^{2},\mu_{R}^{2},\mu_{F}^{2}\right)
+\delta\Delta_{gg}^{A}\left(z,q^{2},\mu_{R}^{2},\mu_{F}^{2}\right)\bigg].
\label{deltaAH}
\end{align}
Eqn. (\ref{deltaAH}) effectively defines $\delta\Delta_{gg}^{A}\left(z,q^{2},\mu_{R}^{2},\mu_{F}^{2}\right)$ as the correction to the scalar Higgs coefficient functions such that the rescaling $g_{0}\left(a_{s}\right)/g_{0}^{H}\left(a_{s}\right)$ converts them to the pseudoscalar coefficients. 
Here, $\Delta_{gg}^{A}\left(z,q^{2},\mu_{R}^{2},\mu_{F}^{2}\right)$ represents the coefficient function for pseudoscalar Higgs boson and $\Delta_{gg}^{H}\left(z,q^{2},\mu_{R}^{2},\mu_{F}^{2}\right)$ represents the same for scalar Higgs boson. 
Moreover, ${g_{0}\left(a_{s}\right)}$ is the constant function of resummation for pseudoscalar Higgs, and ${g_{0}^{H}\left(a_{s}\right)}$ is the analogous function for scalar Higgs. 

\mk{
All the above ingredients are known up to NNLO which has led to the successful computation of $\delta\Delta_{gg}^{A}\left(z,q^{2},\mu_{R}^{2},\mu_{F}^{2}\right)$ up to two-loop level.
It has been shown in \cite{2016} that the $\delta\Delta_{gg}^{A}\left(z,q^{2},\mu_{R}^{2},\mu_{F}^{2}\right)$ corrections vanish at the one-loop level,
and at the two-loop level, these $\delta\Delta_{gg}^{A}\left(z,q^{2},\mu_{R}^{2},\mu_{F}^{2}\right)$ corrections contain only the next-to-next-to-soft terms. 
It is conjectured in \cite{2016} that this can be true to all higher orders.
If that is so, then these $\delta\Delta_{gg}^{A}\left(z,q^{2},\mu_{R}^{2},\mu_{F}^{2}\right)$ corrections do not contain any NSV terms at ${\cal{O}}(a_{s}^{3})$.
Moreover, in \cite{2016} the authors suggest that to define an approximate $\Delta_{gg}^{A}\left(z,q^{2},\mu_{R}^{2},\mu_{F}^{2}\right)$ at N$^3$LO, the unknown ${\cal{O}}(a_{s}^{3})$ contributions to $\delta\Delta_{gg}^{A}\left(z,q^{2},\mu_{R}^{2},\mu_{F}^{2}\right)$ in Eqn. \ref{deltaAH} can be set to zero with a sufficiently good approximation.
By setting $\delta\Delta_{gg}^{A}\left(z,q^{2},\mu_{R}^{2},\mu_{F}^{2}\right)$ to zero this way in Eqn. \ref{deltaAH}, one can obtain the approximate N$^3$LO cross-sections denoted by N$^3$LO$_A$ \cite{Ball_2013,Bonvini:2014joa,Bonvini:2014tea,Bonvini:2015ira,Bonvini_2016_TROLL,Bonvini_2014_TROLL,2016,Bonvini_2018,Bonvini_2018_1,Bonvini:2018xvt}.
Hence, in our analysis we simply rescale the Higgs SV$+$NSV coefficient functions to 
obtain the corresponding ones of the pseudoscalar as 
\begin{align}
\Delta_{gg}^{A,NSV}\left(z,q^{2},\mu_{R}^{2},\mu_{F}^{2}\right)= & \dfrac{g_{0}\left(a_{s}\right)}{g_{0}^{H}\left(a_{s}\right)}\bigg[\Delta_{gg}^{H,NSV}\left(z,q^{2},\mu_{R}^{2},\mu_{F}^{2}\right)
\bigg].
\label{eq:deltaAHnew}
\end{align}
The rescaling components, $g_0(a_s)$ and $g_0^H(a_s)$, are known from resummation 
\cite{Bonvini:2014joa,CATANI1989323,Anastasiou:2014vaa,Laenen:2008ux} while the scalar Higgs coefficient function, $\Delta_{gg}^{H,NSV}\left(z,q^{2},\mu_{R}^{2},\mu_{F}^{2}\right)$, are obtained from \cite{Anastasiou:2014lda,Anastasiou:2016cez}. 
%
The rescaling ratio $g_{0}\left(a_{s}\right)/g_{0}^{H}\left(a_{s}\right)$ up to $a_{s}^{3}$ order is given below:
\begin{align}
\dfrac{g_{0}\left(a_{s}\right)}{g_{0}^{H}\left(a_{s}\right)}= & 1 + a_s (8~C_A) -\frac{1}{3} a_s^2 \bigg[-215~C_A^2+2 ~C_A ~n_f + 3 ~C_F ~n_f \bigg\{ 31-12~
    \log\bigg(\dfrac{m_t^2}{\mu_R^2}\bigg)\bigg\} \bigg] 
    \nonumber \\
     & + \frac{1}{81} a_s^3
\bigg[C_A^3 \bigg(-11880 ~\zeta_2-5616~ \zeta_3+68309\bigg)+C_A^2~
    n_f \bigg(-216 ~\zeta_2-1296 ~\zeta_3+1973\bigg) 
    \nonumber \\
     & + C_A~ C_F~ n_f 
    \bigg\{7776~ \log\bigg(\dfrac{m_t^2}{\mu_R^2}\bigg) -7128 ~\zeta_2+6048
    ~\zeta_3-67094\bigg\} + n_f \bigg(432 ~\zeta_2-631\bigg)
    \nonumber \\
     & + 9~ C_F^2~ n_f \bigg(96 ~\zeta_3+763\bigg)+8~ C_F~ n_f^2 \bigg(162
    ~\zeta_2+565\bigg)-324~C_J^{(2)}\bigg].
 \label{eq:g0Abg0H}
\end{align} 

It has been shown that the $\varphi_{g,i}^{\left(k\right)}$'s given in eqns.~[\ref{eq:phigk1}]-[\ref{eq:phigk3}] for the scalar and the pseudoscalar Higgs boson productions in gluon fusion are identical to each other at two-loop level. The same is also noticed in the case of DY process and scalar Higgs production \textit{via} bottom quark annihilation up to two-loop level. However, for the quark annihilation process it is found that such universality breaks down at third order for $k=0,1$. 
This was checked in \cite{ajjath2020soft} using the state-of-art results from \cite{Anastasiou_2015,Duhr:2019kwi,Mistlberger:2018etf,Duhr:2020seh}. 
In our present work, when we explicitly compute $\varphi_{g,3}^{\left(0\right)}$ and $\varphi_{g,3}^{\left(1\right)}$ from Eqn.~\ref{eq:deltaAHnew}, we notice that they are same for both scalar and pseudoscalar Higgs boson productions {\it{via}} gluon fusion.
Hence, the universality of the $\varphi_{g,i}^{\left(k\right)}$'s at third order can be checked only when the explicit N$^3$LO results are available for the pseudoscalar Higgs boson production in gluon fusion.
}

As depicted in Section 3 of \cite{ajjath2020soft}, once we have the $\varphi_{g,i}^{\left(k\right)}$ values up to a certain order, we can predict the coefficients of the highest logarithms in the coefficient functions to a few consecutive higher orders of $\Delta_{gg}^{A,NSV}\left(z,q^{2},\mu_{R}^{2},\mu_{F}^{2}\right)$ computed using the formalism developed in \cite{Ravindran_2006,Ravindran_2006_B,ajjath2020soft}.
Using the evaluated $\varphi_{g,i}^{\left(k\right)}$ values, we could predict the three highest logarithms of $\Delta_{gg}^{A,NSV}\left(z,q^{2},\mu_{R}^{2},\mu_{F}^{2}\right)$, up to ${\cal{O}}(a_{s}^{7})$ from ${\cal{O}}(a_{s}^{4})$. 
\mk{ 
Even when we use the available $\varphi_{g,i}^{\left(k\right)}$'s for $k=0,1$, we could predict these highest logarithmic terms because they are independent of $\varphi_{g,3}^{\left(0\right)}$ and $\varphi_{g,3}^{\left(1\right)}$.
}
The ingredients needed to exactly match the other higher order results of the pseudoscalar Higgs boson are still not available. 
So, we compute the highest logarithms up to ${\cal{O}}(a_{s}^{7})$ from ${\cal{O}}(a_{s}^{4})$ of the $\Delta_{gg}^{A,NSV}\left(z,q^{2},\mu_{R}^{2},\mu_{F}^{2}\right)$ result. 
%

The explicit results of the SV+NSV cross-section up to N$^3$LO are given in the Appendix \ref{appendix:A} where it can be seen that the general form of the output is depicted as:
\begin{equation}
\Delta_{g,i}^{A,NSV}\left(z,q^{2}\right)=\delta(1-z)[\cdots] + \sum_{j=1/2,1,\cdots}^{i}\mathcal{D}_{(2j-1)}[\cdots] + \sum_{j=1/2,1,\cdots}^{i}\log^{(2j-1)}(1-z)[\cdots].
\end{equation}
where $i = 1,2,3$, and $[\cdots]$ represent coefficients of the corresponding distribution, logarithmic term or delta function. 
The constant terms in the above result actually correspond to the $\log^0(1-z)$ coefficients.

\section{Resummation of the NSV results in Mellin Space}
\label{sec:resNSVformalism}

In order to study the all order behavior of the coefficient function, $\Delta_{gg}$, in the $N$-moment space, it is convenient to use the following form of the partonic coefficient function \cite{ajjath2020soft}:   
\begin{eqnarray}
\label{resumz}
\Delta_{gg}(q^2,\mu_R^2,\mu_F^2,z)= C^g_0(q^2,\mu_R^2,\mu_F^2) 
~~{\cal C} \exp \Bigg(2 \Psi^g_{\cal D} (q^2,\mu_F^2,z) \Bigg)\,,
\end{eqnarray}
where
\begin{eqnarray}
\label{phicint}
\Psi^g_{\cal D} (q^2,\mu_F^2,z) &=& {1 \over 2}
\int_{\mu_F^2}^{q^2 (1-z)^2} {d \lambda^2 \over \lambda^2} 
P_{gg} (a_s(\lambda^2),z)  + {\cal Q}^g(a_s(q^2 (1-z)^2),z)\,,
\end{eqnarray}
with
\begin{eqnarray}
\label{calQc}
{\cal Q}^g (a_s(q^2(1-z)^2),z) &=&  \left({1 \over 1-z} \overline G^g_{SV}(a_s(q^2 (1-z)^2))\right)_+ + \varphi_{f,g}(a_s(q^2(1-z)^2),z).
\end{eqnarray}
The coefficient $C_0^g$ is $z$ independent and is expanded in powers of $a_s(\mu_R^2)$ as
\begin{eqnarray}
\label{C0expand}
C_0^g (q^2,\mu_R^2,\mu_F^2) = \sum_{i=0}^\infty a_s^i(\mu_R^2) C_{0i}^g (q^2,\mu_R^2,\mu_F^2)\,,
\end{eqnarray}
where the coefficients $C^g_{0i}$ are calculated in \cite{Ahmed_2015} for pseudoscalar. Eqn.(\ref{resumz}) gives the $z$-space resummed result.

Now it is easy to compute the Mellin moment of $\Delta_{gg}$. 
The limit $z\rightarrow 1$ translates to $N\rightarrow \infty$ in the $N$-moment space and to include NSV terms, we need to keep ${\cal O}(1/N)$ corrections in the large $N$ limit.  
The Mellin moment of $\Delta_{gg}$ is given by \cite{ajjath2020soft}
\begin{eqnarray}
\label{DeltaN}
\Delta_{gg,N}(q^2,\mu_R^2,\mu_F^2) = C_0(q^2,\mu_R^2,\mu_F^2) \exp\left(
\Psi_N^g (q^2,\mu_F^2) 
\right)\,,
\end{eqnarray}
The Mellin moment of the exponent acquires the following form: 
\begin{eqnarray}
\Psi_N^g = \Psi_{\rm{SV},N}^g + \Psi_{\rm{NSV},N}^g ,
\end{eqnarray}
where we can split $\Psi_N^g$ in such a way that all those terms that are functions of $\log^j(N),~j=0,1,\cdots$ are
kept in $\Psi_{{\rm SV},N}^g$ and the remaining terms that are proportional to $(1/N) \log^j(N),~j=0,1,\cdots$ are contained
in $\Psi_{\rm{NSV},N}^g$.  Hence,
\begin{eqnarray}
\label{PsiSVN}
	\Psi_{\rm{SV},N}^g = \log(g_0^g(a_s(\mu_R^2))) + g_1^g(\omega)\log(N) + \sum_{i=0}^\infty a_s^i(\mu_R^2) g_{i+2}^g(\omega) \,,
\end{eqnarray}
where the $g^g_i(\omega)$'s, identical to those in \cite{CATANI1989323,Moch:2005ba,H:2019dcl}, are obtained from the resummed formula for SV terms
and $g^g_0(a_s)$ is expanded in powers of $a_s$ as (see \cite{Moch:2005ba})
\begin{eqnarray}
	\log(g_0^g(a_s(\mu_R^2))) = \sum_{i=1}^\infty a_s^i(\mu_R^2) g^g_{0,i}\quad \,.
\end{eqnarray}
The $g_i^g(a_s(\mu_R^2))$'s are also provided in the ancillary files of \cite{ajjath2020soft}.

The function $\Psi_{\rm{NSV},N}^g$ is given by
\begin{align}
\label{PsiNSVN}
 \Psi_{\rm{NSV},N}^g = {1 \over N} \sum_{i=0}^\infty a_s^i(\mu_R^2) \bigg( 
 \bar g_{i+1}^g(\omega)  +  h^g_{i}(\omega,N) \bigg)\,,
\end{align}
with 
\begin{align}
	h^g_i(\omega,N) = \sum_{k=0}^{i} h^g_{ik}(\omega)~ \log^k(N).
\label{eq:hcik}	
\end{align}
where the resummation constants, $\bar g_{i}^g(\omega) $ and $h^g_i(\omega,N)$, are given in \cite{ajjath2020soft}.

\section{Numerical Results and Discussion}
\label{sec:Numerical}
In this section, we will present our numerical results of the NSV corrections at N$^3$LO level in QCD for the production of a pseudoscalar Higgs boson at the LHC. 
Our predictions are based on EFT where the top quarks are integrated out at higher orders. However, we retain the top quark mass dependence at LO.
The term C$_J^{(2)}$ in the Wilson coefficient $C_J(a_s)$ is taken to be zero in our analysis because it is not available in the literature yet.  
For simplicity, we have set cot $\beta=1$ in our numerical analysis. Results for other values of cot $\beta$ can be easily obtained by rescaling the cross-sections with cot$^2\beta$.
At LO, we have retained the full top quark mass dependence while EFT approach has been used for higher order corrections.
We use MMHT 2014 PDFs throughout where the LO, NLO and NNLO parton level cross-sections are 
convoluted with the corresponding order by order central PDF sets, but for N$^3$LO cross-sections we simply use MMHT2014nnlo68cl PDFs. 
The strong coupling constant is provided by the respective PDFs from LHAPDF.

To estimate the impact of QCD corrections, we define the K-factors as
\begin{equation}
K_{\left(1\right)}^X=\dfrac{\sigma_\text{NLO}^X}{\sigma_\text{LO}},\ \ \ \ \ K_{\left(2\right)}^X=\dfrac{\sigma_\text{NNLO}^X}{\sigma^\text{LO}},
\label{eq:KfactorFO}
\end{equation}
where X is either SV or SV+NSV or Full which includes all possible sub-processes or Full$(gg)$ which includes only the gluon-gluon$(gg)$ sub-process. 
When the Full$(gg)$ case is computed, only the $gg$ sub-process is considered at the $n$-th order while at the lower orders $(k<n)$, all possible sub-processes are taken together.

\begin{figure}[!htb]
\centering
\begin{subfigure}{.5\textwidth}
  \centering
  \includegraphics[width=1.0\linewidth]{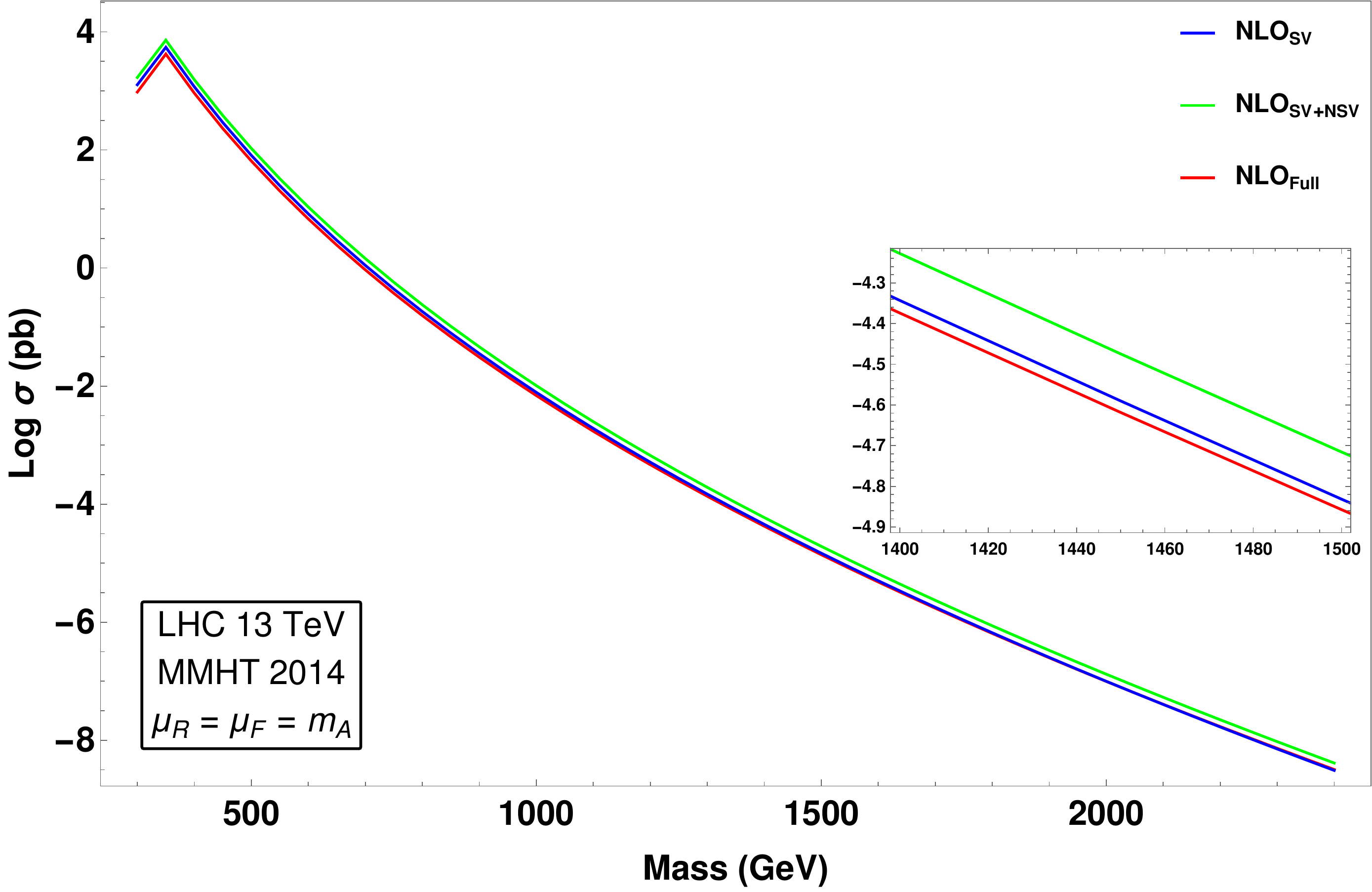}
  \label{fig:sub1}
\end{subfigure}%
\begin{subfigure}{.5\textwidth}
  \centering
  \includegraphics[width=1.0\linewidth]{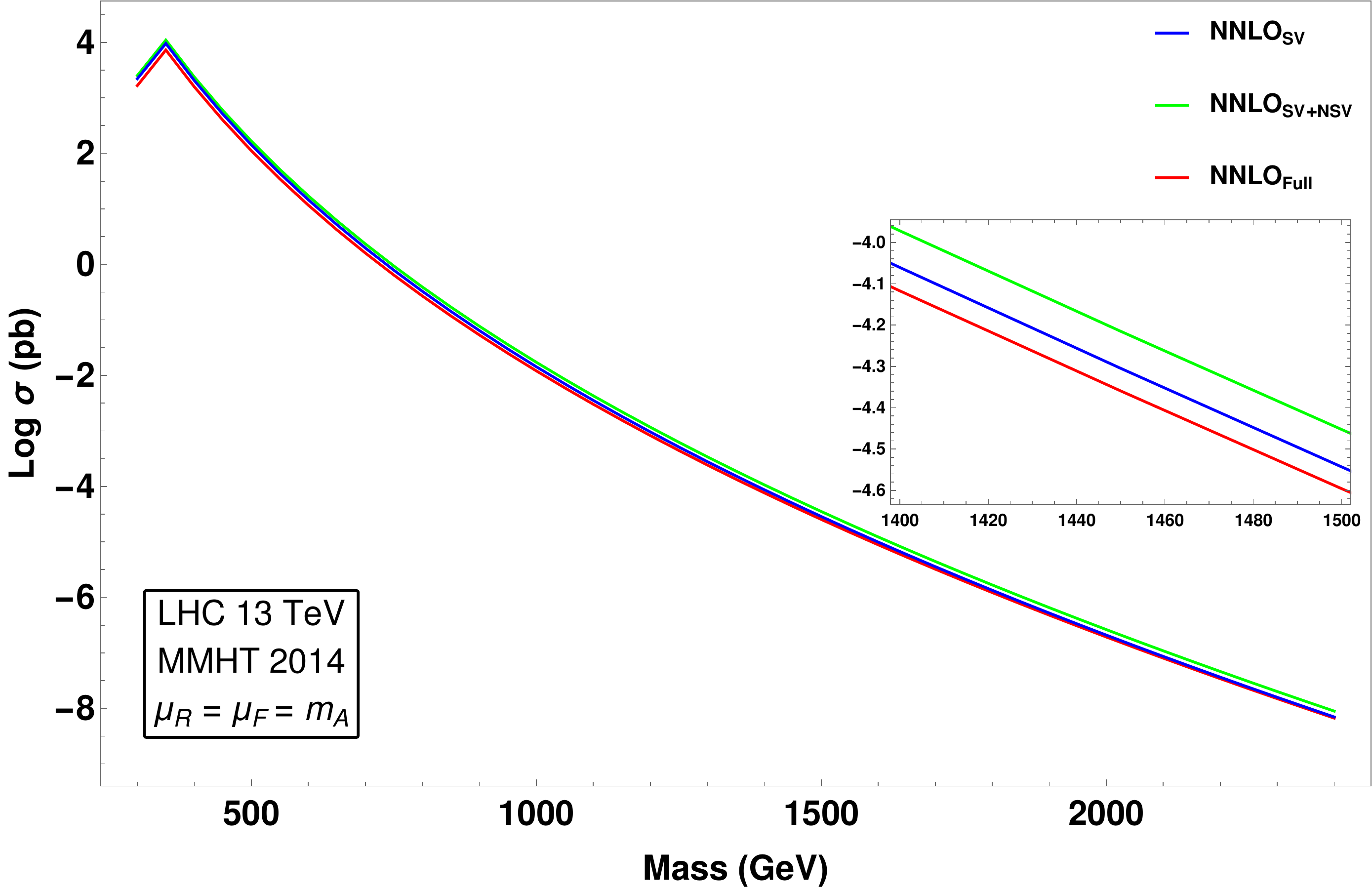}
  \label{fig:sub2}
\end{subfigure}
\caption{Pseudoscalar production cross-section at NLO level (left panel) and NNLO level (right panel) with a comparison between fixed order SV, SV+NSV and Full results for 13 TeV LHC}
\label{fig:cross12}
\end{figure}
\begin{figure}[!htb]
\centering
\begin{subfigure}{.5\textwidth}
  \centering
  \includegraphics[width=1.0\linewidth]{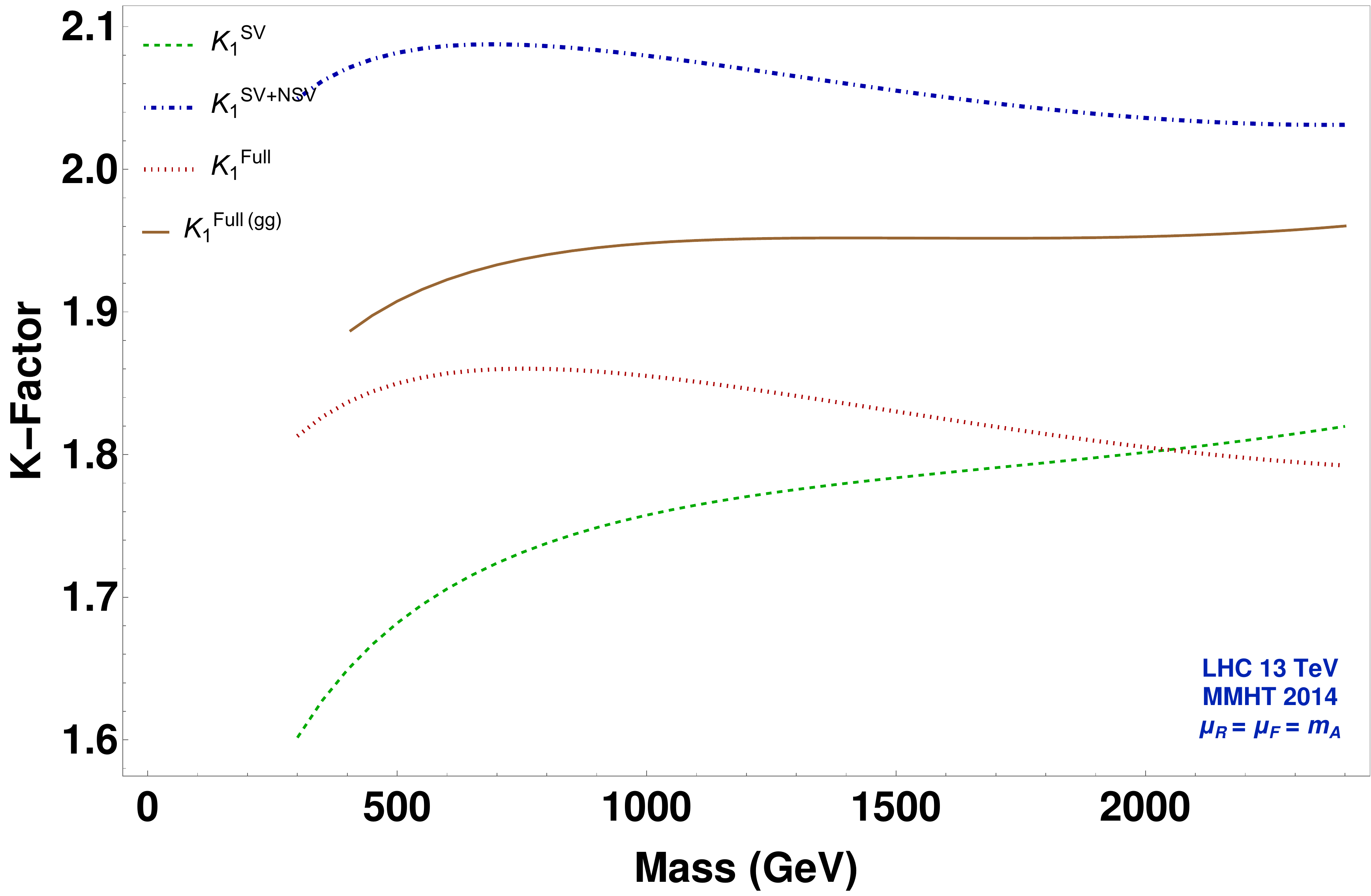}
  \label{fig:sub1}
\end{subfigure}%
\begin{subfigure}{.5\textwidth}
  \centering
  \includegraphics[width=1.0\linewidth]{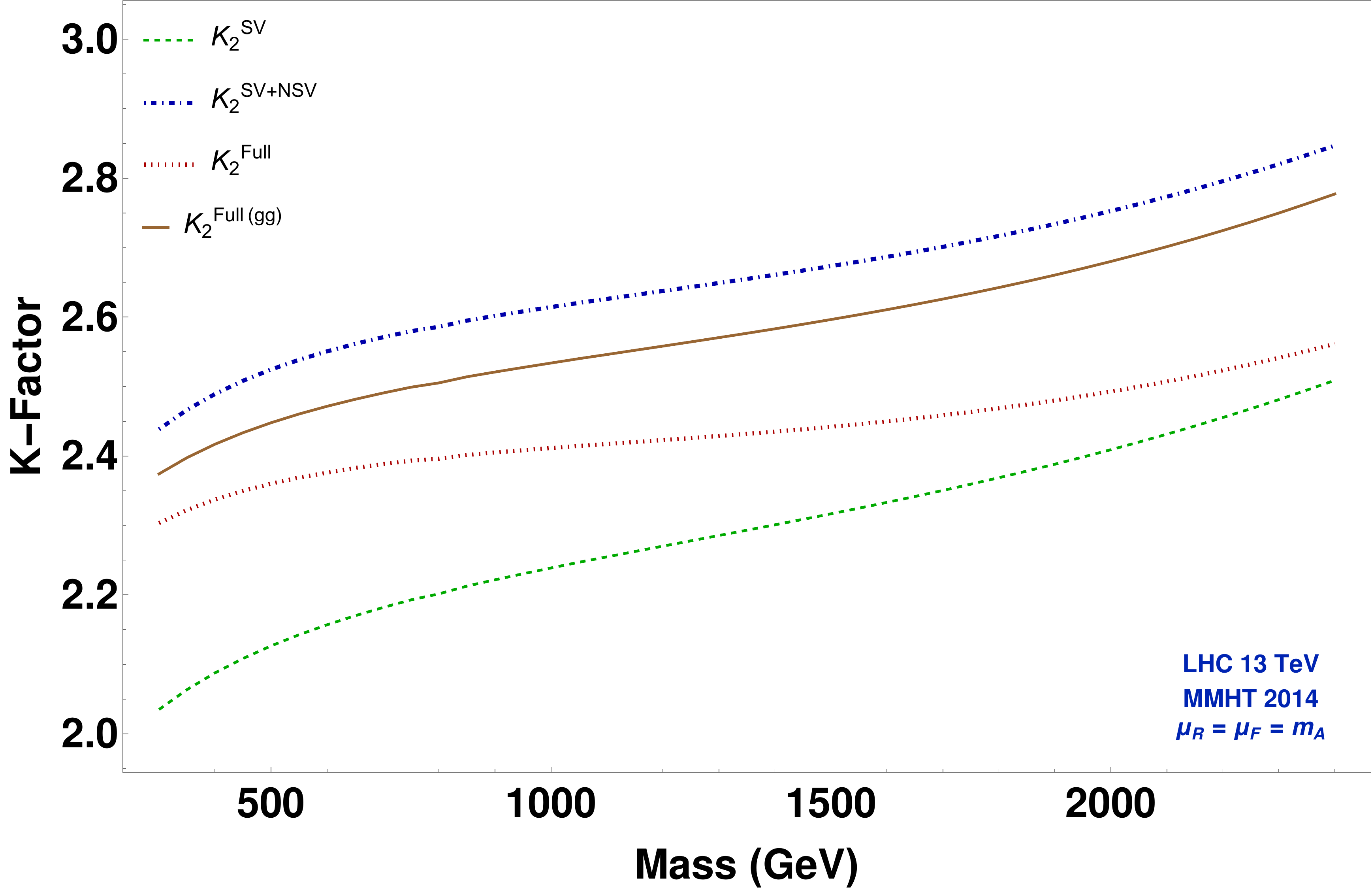}
  \label{fig:sub2}
\end{subfigure}
\caption{K-Factor for pseudoscalar production cross-section at NLO level (left panel) and NNLO level (right panel) with a comparison between fixed order SV, SV+NSV,  Full (all sub-processes included) and Full (only gg sub-process included at $n$-th order) results for 13 TeV LHC.}
\label{fig:kfactor12}
\end{figure}

In Fig.~\ref{fig:cross12}, we plot the pseudoscalar production cross-section as a function of its mass m$_A$ at NLO(left panel) and NNLO (right panel), by varying $m_A$ from $300$ GeV to $2400$ GeV.  
Here the ``SV'' results contain the SV threshold logarithms and $\delta(1-z)$ contributions, while the ``NSV'' corrections include the NSV logarithms in the gluon fusion channel only. 
We note that the NSV corrections at higher orders also arise from other partonic channels, a detailed study of which is beyond the scope of the present work and will be 
presented elsewhere.  
As can be seen from fig.~\ref{fig:cross12}, the SV results give sizable contribution to the fixed order results however they underestimate the latter. The inclusion of NSV corrections increases the cross-section substantially but they overestimate the FO results for the mass range of $m_A$ we have considered. 
%
%

To study the impact of these corrections better at SV and beyond, we depict in Fig.~\ref{fig:kfactor12} the corresponding K-factors defined in eqn. {\ref{eq:KfactorFO}}.  
Since our NSV corrections include only the gluon fusion channel, it is worthwhile to have comparison with the complete result of the gluon fusion channel including the pure regular contributions. 
From these K-factors, we notice that the SV corrections converge to the FO result (denoted by $K_i^{Full}$) in the high mass region and differ significantly in the small mass region. However, the FO result here also does contain contributions from other parton channels, and the difference noticed between the full FO result and the pure gluon channel (denoted by $K_i^{Full(gg)}$) contribution can be understood from the presence of these other channels.  
We also notice that the SV+NSV corrections have a similar behavior to that of the complete
gluon fusion channel contribution, with the former being a little bit higher than the latter. 
\mk{This also indicates that the regular or beyond NSV corrections have a negative impact but are smaller in magnitude compared to the NSV corrections at higher orders.}


Next, we study the impact of resummation of these NSV logarithms on the pseudoscalar production cross-section at NLL and NNLL accuracy. 
For the resummed cross-section, we do the matching as below:
\begin{equation}
 \sigma^{(\text{matched})}=\sigma^{\text{SV+NSV}}_{\text{resum}}-\sigma^{\text{SV+NSV}}\bigg|_{(\text{FO})}+\sigma^{(\text{FO})}.
\end{equation}

\begin{figure}[!htb]
\centering
  \includegraphics[keepaspectratio,width=0.75\linewidth]{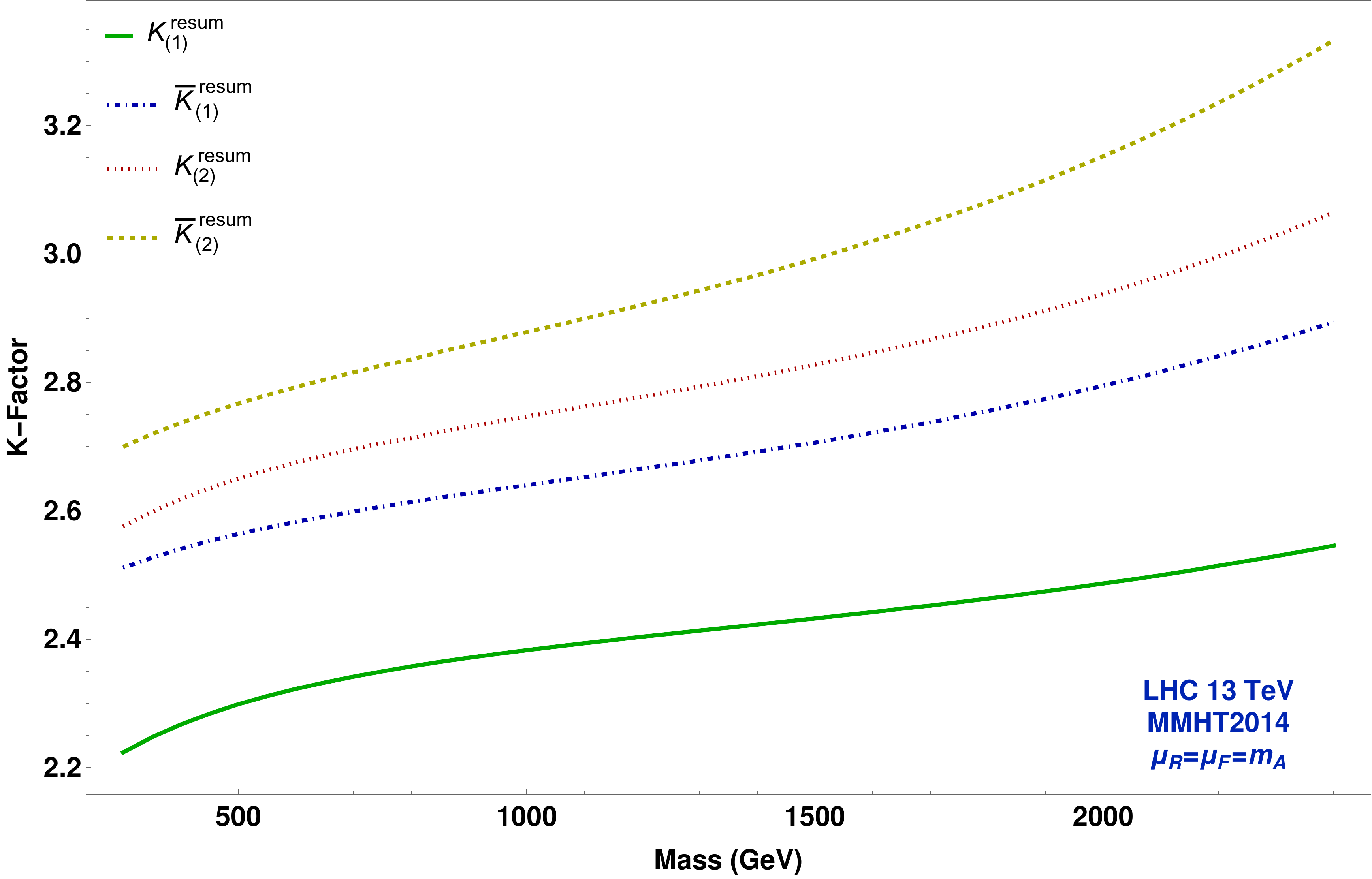}
  \caption{Resummed K-factor plots for 13 TeV LHC taking MMHT 2014 as the reference PDF and choosing the central scale $\mu_R = \mu_F = m_A$.}
  \label{fig:sub1}
\label{fig:resumKfactor}
\end{figure}

In Fig.~\ref{fig:resumKfactor}, we depict the resummed K-factors at NLO and NNLO orders, and contrast them against the corresponding ones due to the SV resummation for a wide range of pseudoscalar mass {\it{i.e.}} $300 < m_A < 2400$ GeV. 
We define these K-factors as
\begin{align}
K_{\left(1\right)}^\text{resum}=\dfrac{\sigma_\text{NLO+NLL}}{\sigma_\text{LO}},\ \ \ \ \ K_{\left(2\right)}^\text{resum}=\dfrac{\sigma_\text{NNLO+NNLL}}{\sigma^\text{LO}},
\nonumber \\
\nonumber \\
\overline{K}_{\left(1\right)}^\text{~resum}=\dfrac{\sigma_{\text{NLO}+\overline{\text{NLL}}}}{\sigma_\text{LO}},\ \ \ \ \ \overline{K}_{\left(2\right)}^\text{~resum}=\dfrac{\sigma_{\text{NNLO}+\overline{\text{NNLL}}}}{\sigma^\text{LO}},
\end{align}
The resummed NLO SV K-factor $({K}_{(1)}^{\text{resum}})$ varies from 2.2 (at $m_A=300$ GeV) to about 2.5 (at $m_A=2400$ GeV).
However, the inclusion of NSV logarithms at $\overline{\text{NLL}}$ accuracy increase these results by about 30\% in the low mass region, and by about 40\% of LO in the high mass region {\it{i.e.}} $(\overline{K}_{(1)}^{~\text{resum}})$ varies from $2.5$ to about $2.9$.
The resummation of NSV logarithms to $\overline{\text{NNLL}}$ accuracy has a similar behavior and enhance the SV resummed results by about 
\mk{10\% (30\%)} 
in the low (high) mass region.

%
%
%

\begin{figure}[!htb]
\centering
\subfloat{%
  \includegraphics[clip,width=0.7\columnwidth]{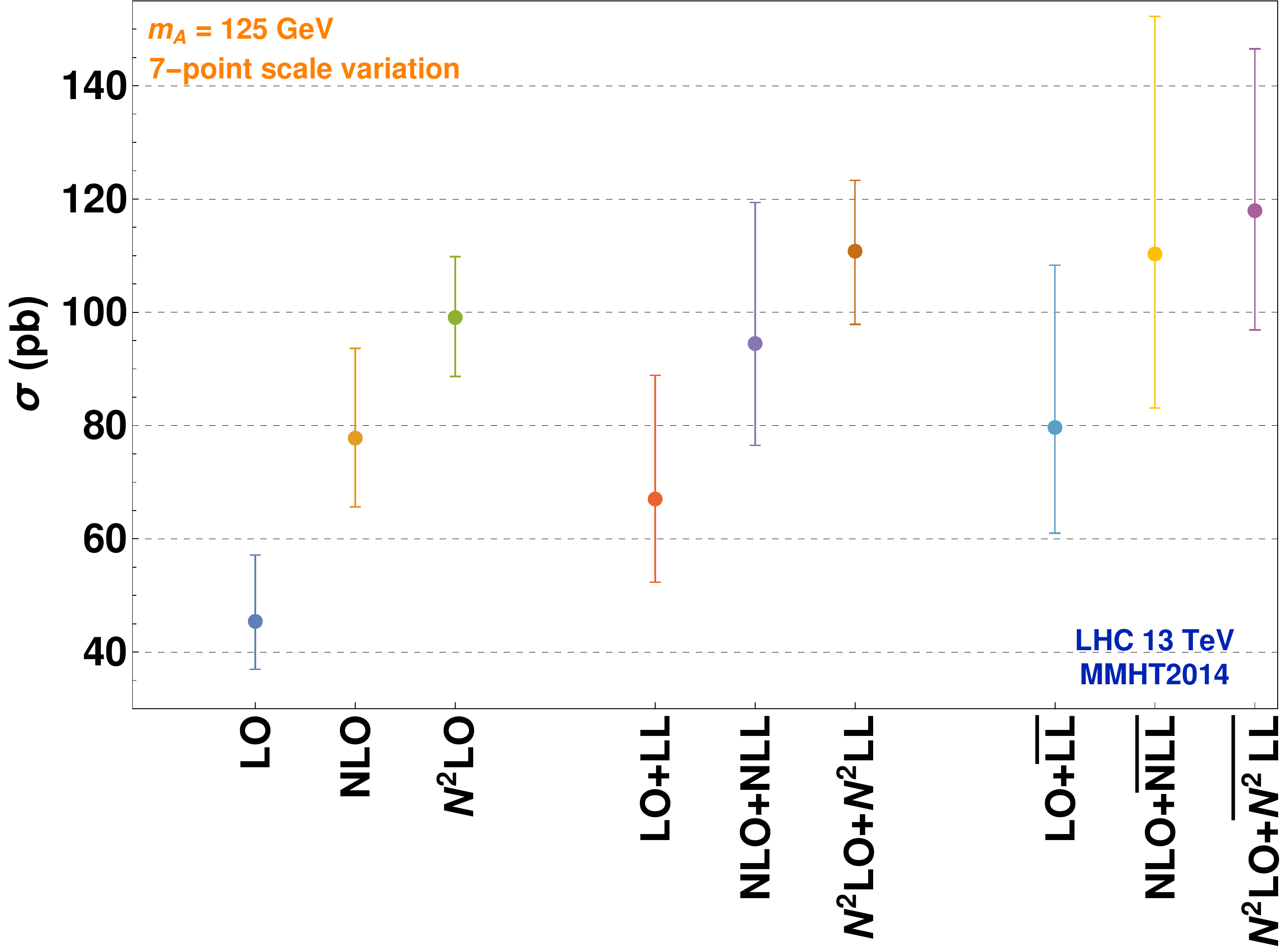}%
}

\subfloat{%
  \includegraphics[clip,width=0.7\columnwidth]{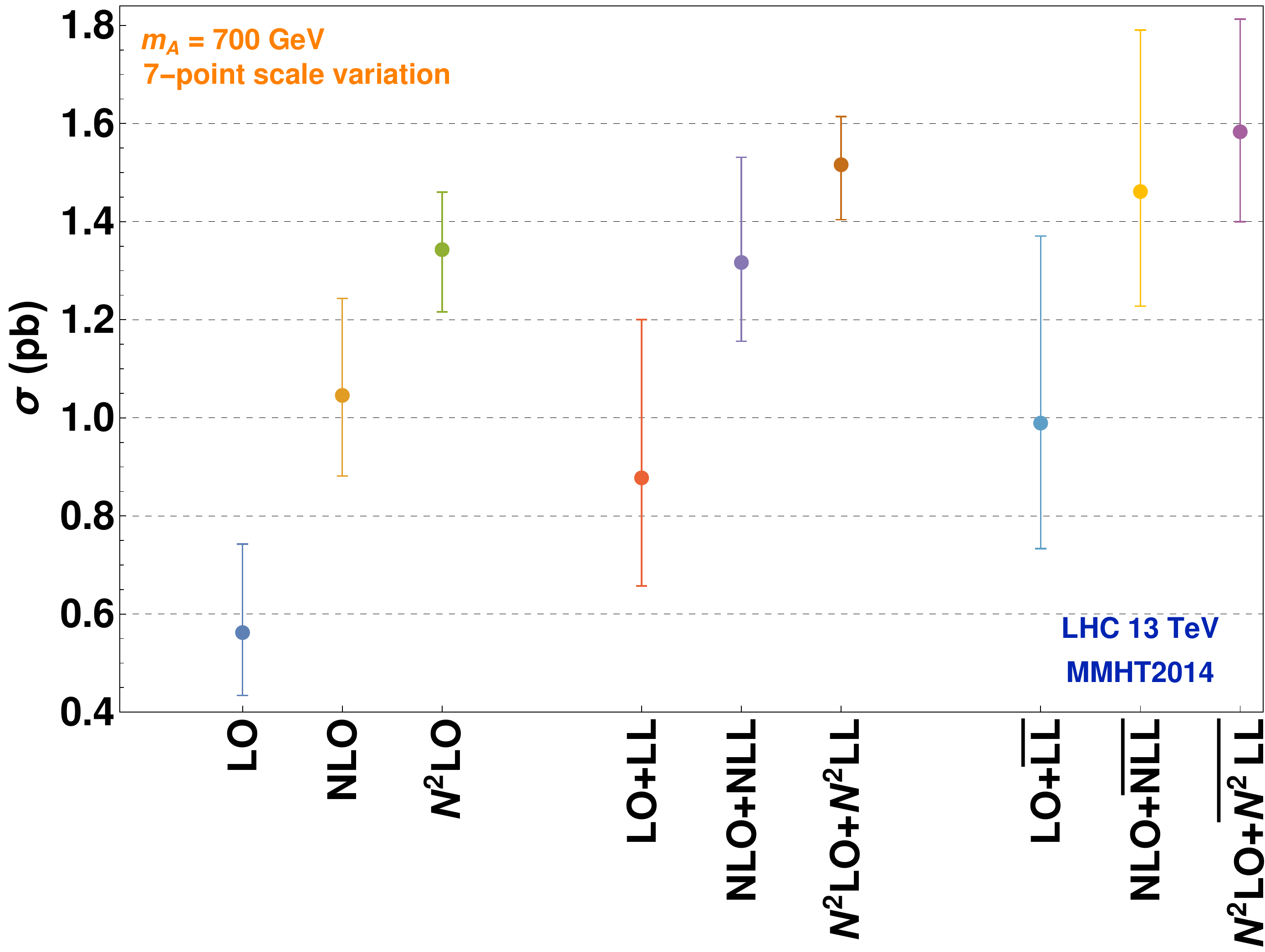}%
}
\caption{
Uncertainty plot with 7-point scale uncertainty for $m_A$=125 GeV (top figure) and $m_A$=700 GeV (bottom figure) for 13 TeV LHC with MMHT 2014 PDF.}
\label{fig:PS_7pt_ErrorBar}
\end{figure}

\begin{figure}[!htb]
\centering
\subfloat{%
  \includegraphics[clip,width=0.7\columnwidth]{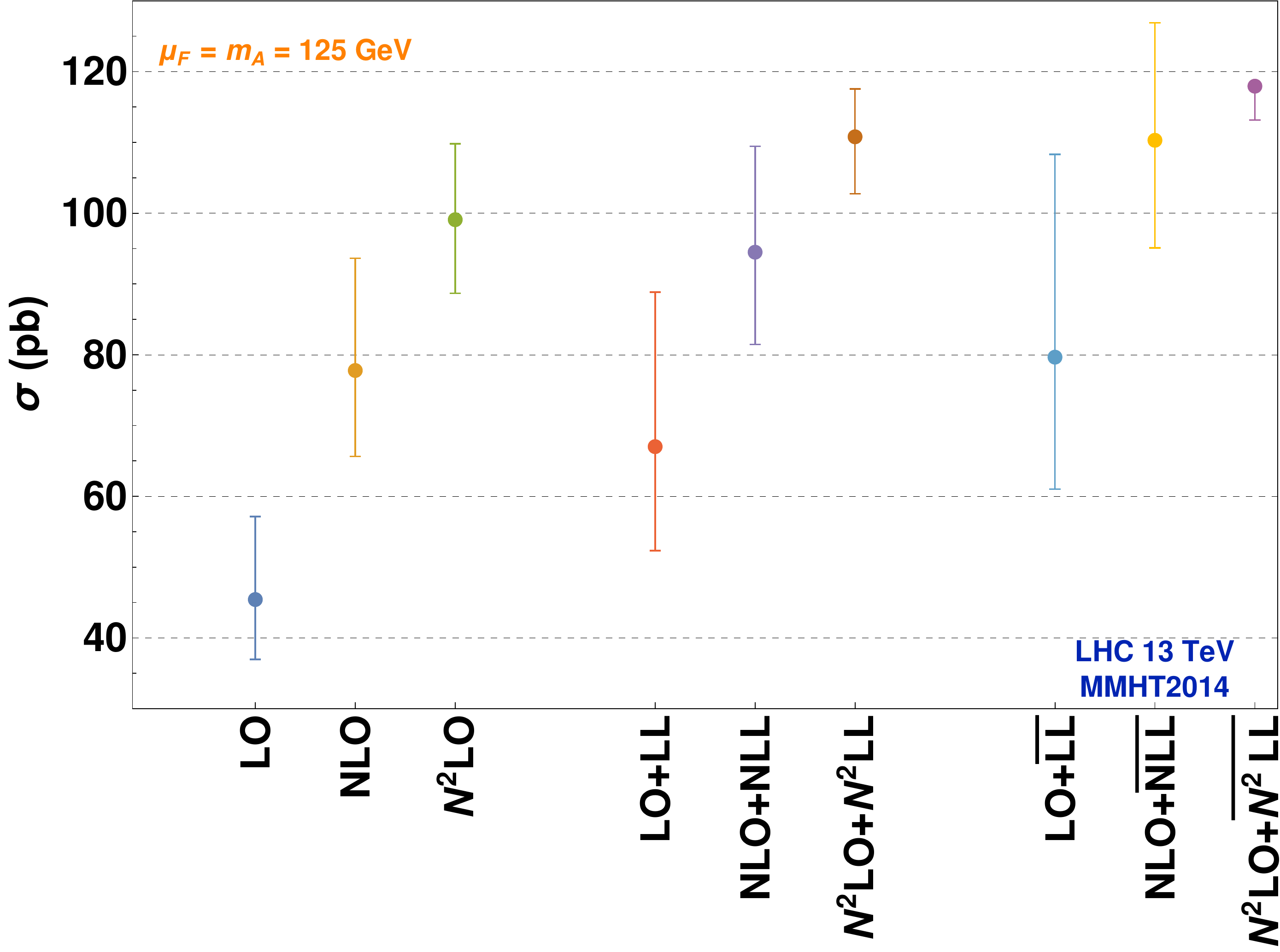}%
}

\subfloat{%
  \includegraphics[clip,width=0.7\columnwidth]{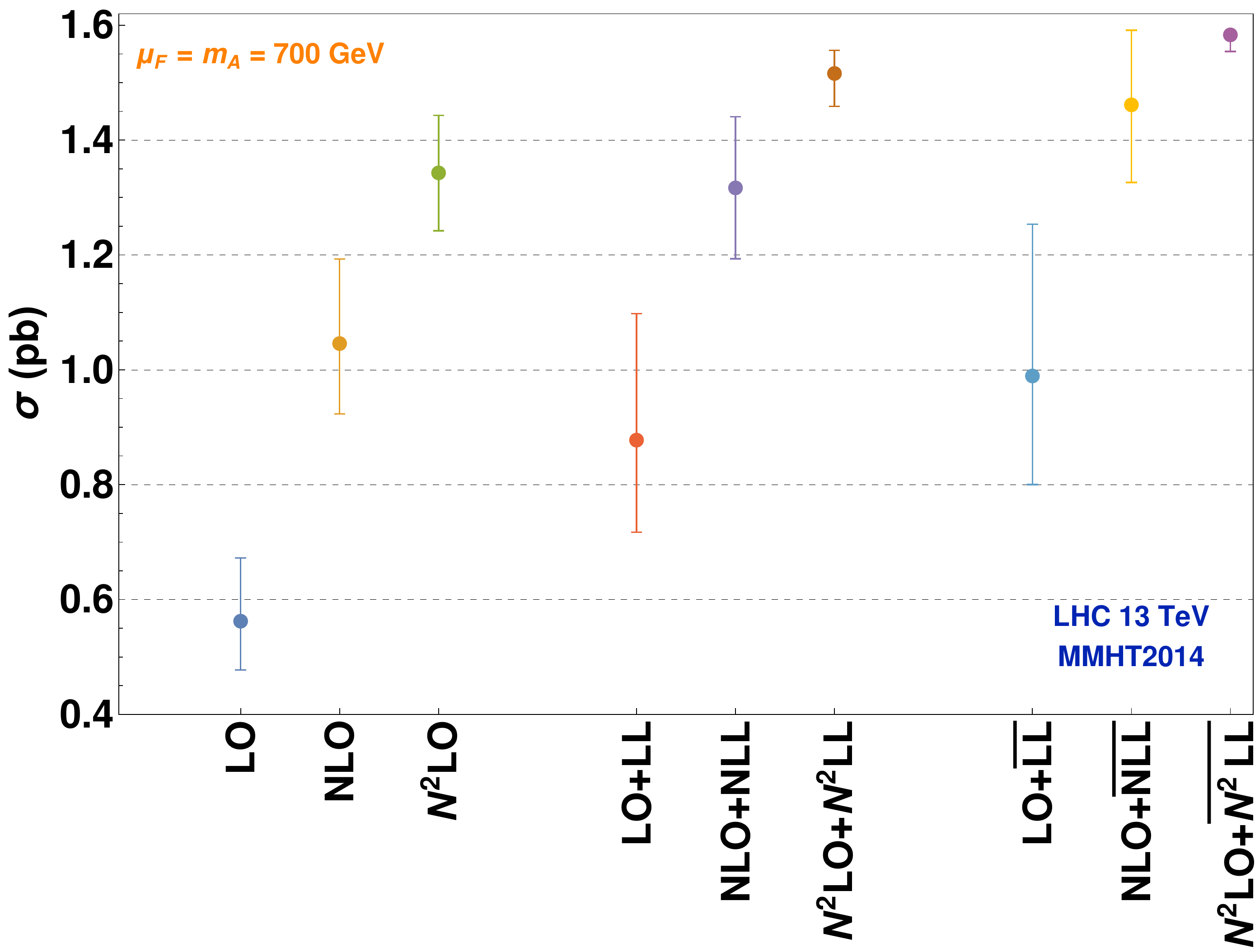}%
}

\caption{Uncertainty plot with $\mu_F$ scale fixed for $m_A$=125 GeV (top figure) and $m_A$=700 GeV (bottom figure) for 13 TeV LHC with MMHT 2014 PDF.}
\label{fig:PS_125GeV_ErrorBar}
\end{figure}

\begin{figure}[!htb]
\centering
\subfloat{%
  \includegraphics[clip,width=0.7\columnwidth]{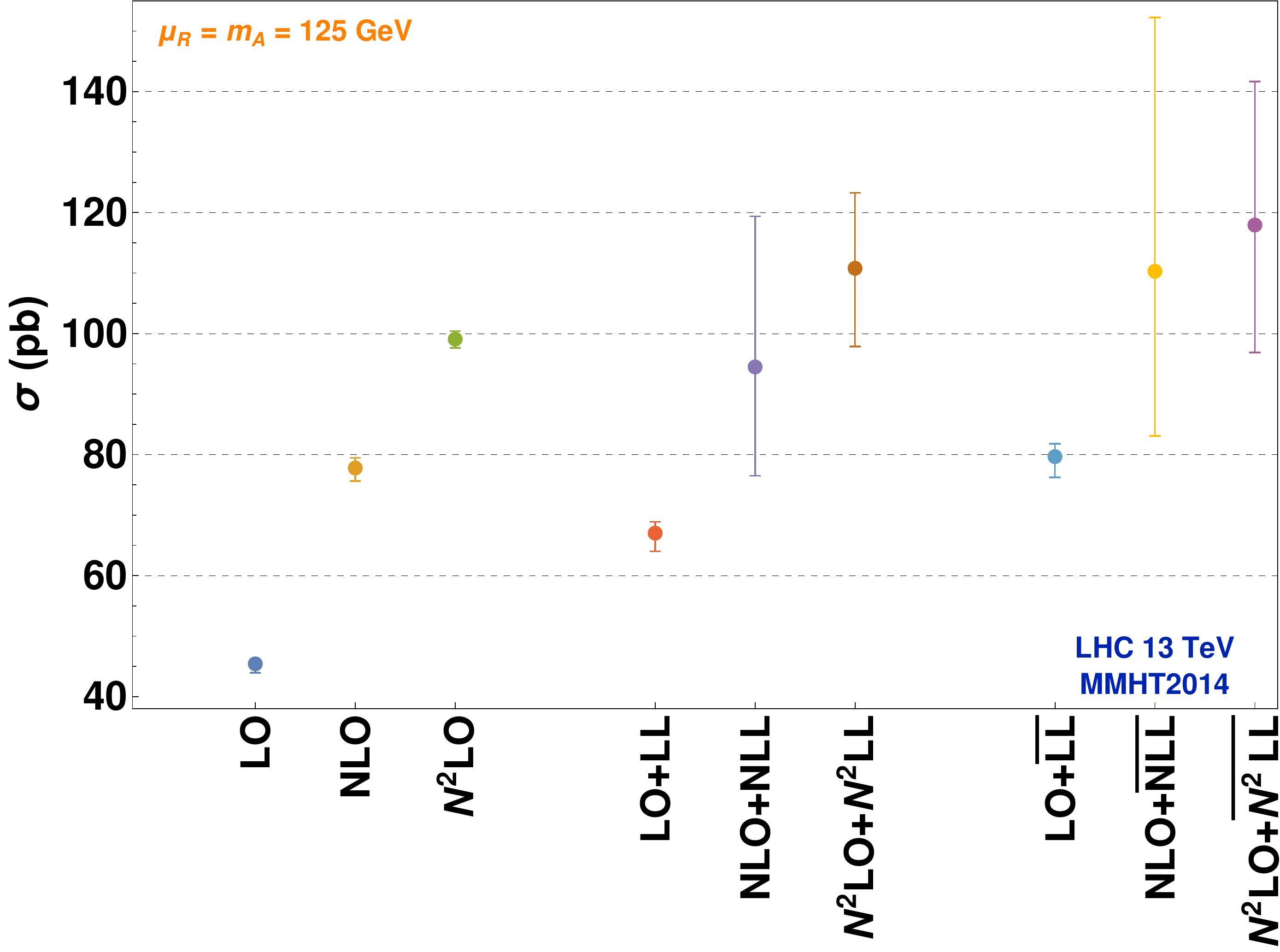}%
}

\subfloat{%
  \includegraphics[clip,width=0.7\columnwidth]{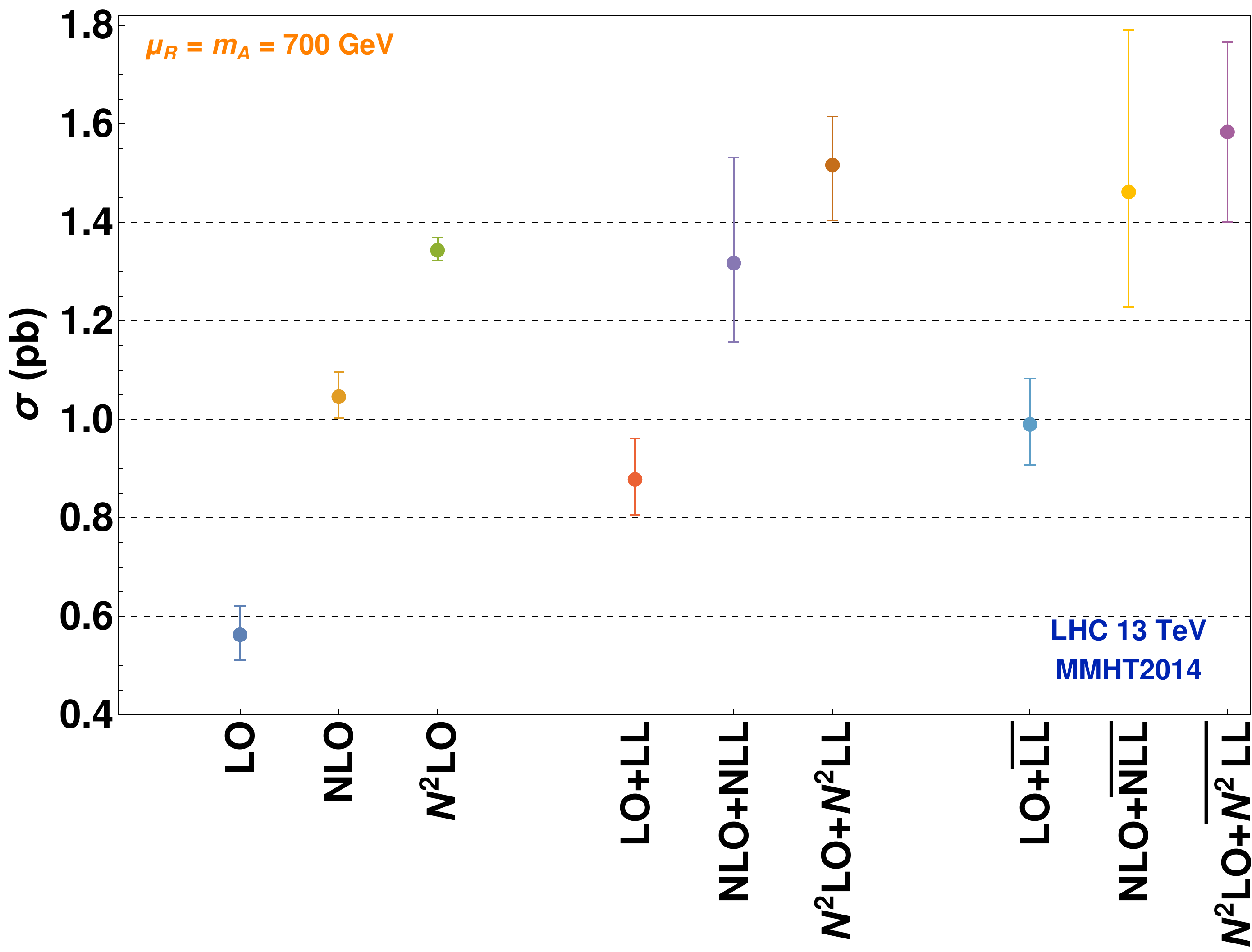}%
}  
  
\caption{Uncertainty plot with $\mu_R$ scale fixed for $m_A$=125 GeV (top figure) and $m_A$=700 GeV (bottom figure) for 13 TeV LHC with MMHT 2014 PDF.}
\label{fig:PS_700GeV_ErrorBar}
\end{figure}

We will next study the theoretical uncertainties due to the unphysical scales, $\mu_R$ and $\mu_F$, in our predictions.
We will present the conventional $7$-point scale uncertainties and make the following scale choices 
($\mu_R/m_A, \mu_F/m_A)$: $(0.5,0.5), (0.5,1), (1.0, 0.5), (1.0, 1.0)$, 
$(1.0,2.0), (2.0,1.0)$ and $(2.0,2.0)$ for a given value of $m_A$.
\mk{ 
All these uncertainties are presented to NNLO+NNLL and NNLO+$\overline{\text{NNLL}}$ accuracy.
In all the uncertainty plots, the first three results correspond to FO results, the next three correspond to SV resummation and the last three represent the NSV resummed ones. 
We first plot the $7$-point scale uncertainty involved for the total pseudoscalar production cross-section at each order for m$_A=125$ GeV (top) and m$_A=700$ GeV (bottom) in Fig. \ref{fig:PS_7pt_ErrorBar} . 
We observe in Fig. \ref{fig:PS_7pt_ErrorBar} that the 7-point scale uncertainties get reduced on going from NLO to NNLO, NLO+NLL to NNLO+NNLL and NNLO+$\overline{\text{NNLL}}$ for both m$_A=125$ and $700$ GeV. 
However, this 7-point scale uncertainties are found to increase while going from the SV to NSV resummation. 
To better understand this aspect, we study the scale variations 
due to $\mu_R$ and $\mu_F$ separately by varying one of them between $[m_A/2,~2m_A]$ and keeping the other fixed at $m_A$, for both low and high mass regions.
In Fig. \ref{fig:PS_125GeV_ErrorBar}, we present the $\mu_R$ scale uncertainties by keeping $\mu_F$ fixed at m$_A$. 
Here we notice that at second order, the NSV resummed results have smaller scale uncertainties compared to the SV resummed ones which in turn are smaller than the FO scale uncertainties for both m$_A=125$ and $700$ GeV.  
In Fig. \ref{fig:PS_700GeV_ErrorBar}, we study the $\mu_F$ scale uncertainties by keeping $\mu_R$ fixed at m$_A$.  
Here we notice that, contrary to the case of $\mu_R$ scale uncertainties, the NSV resummed results have larger scale uncertainties compared to the SV resummed ones which are in turn larger than the corresponding FO ones for both m$_A=125$ and $700$ GeV.
This can be attributed to the missing contributions from other partonic channels while estimating the $\mu_F$ scale uncertainties in our SV and NSV resummed predictions.
}

%
%
%
%
%
%
%
%
%
%
%

\mk{
\begin{table}[!htb]
\centering
\begin{tabularx}{1.0\textwidth}{ 
  | >{\raggedright\arraybackslash}X
  | >{\centering\arraybackslash}X
  | >{\centering\arraybackslash}X
  | >{\centering\arraybackslash}X
  | >{\centering\arraybackslash}X
  | >{\centering\arraybackslash}X
  | >{\centering\arraybackslash}X
  | >{\centering\arraybackslash}X
  | >{\centering\arraybackslash}X
  | >{\raggedleft\arraybackslash}X | }
 \hline
 \textbf{Order}            & \textbf{Percentage uncertainty (7-point variation)} & \textbf{Percentage uncertainty ($\mu_F$ fixed)}  & \textbf{Percentage uncertainty ($\mu_R$ fixed)} \\ \hline \hline
\rule{0pt}{10pt} LO                            & 54.92 & 34.70 & 19.54 \\ \hline
\rule{0pt}{10pt} NLO                           & 34.61 & 25.81 &  8.91 \\ \hline
\rule{0pt}{10pt} NNLO                          & 18.18 & 14.96 &  3.48 \\ \hline
                                                      \hline   
\rule{0pt}{10pt} LO+LL                         & 61.84 & 43.35 & 17.72 \\ \hline
\rule{0pt}{10pt} NLO+NLL                       & 28.45 & 18.80 & 28.45 \\ \hline
\rule{0pt}{10pt} NNLO+NNLL                     & 13.89 &  6.42 & 13.89 \\ \hline
                                                      \hline
\rule{0pt}{10pt} LO+$\overline{\text{LL}}$     & 64.40 & 45.83 & 17.71 \\ \hline
\rule{0pt}{10pt} NLO+$\overline{\text{NLL}}$   & 38.52 & 18.13 & 38.52 \\ \hline
\rule{0pt}{10pt} NNLO+$\overline{\text{NNLL}}$ & 26.09 &  1.82 & 23.13 \\ \hline
\end{tabularx}
\caption{\mk{Percentage of uncertainty in the pseudoscalar production cross-sections due to scale variations, for the pseudoscalar mass $m_A=700$ GeV at 13 TeV LHC, using MMHT 2014 PDF. The second column corresponds to $7$-point scale variations,
the third column represents the uncertainty due to the $\mu_R$ scale variations while keeping $\mu_F$ fixed and
the last column represents the uncertainty due to the $\mu_F$ scale variations while keeping $\mu_R$ fixed.}}
\label{tab:700GeV_table}
\end{table}
In table~\ref{tab:700GeV_table}, we tabulate the percentage errors due to the 7-point scale uncertainty, $\mu_F$ scale uncertainty and $\mu_R$ scale uncertainty cases for the pseudoscalar mass of $700$ GeV.
To study the effect of parton fluxes on the scale uncertainties in different kinematic regions, we present in table~\ref{tab:7pt_Comp_table} the 7-point scale uncertainties for different m$_A$ values. 
Here we observe that at the NLO and NNLO levels, the 7-point scale uncertainties decrease as we go from 125 GeV to about 1000 GeV and then, they slowly increase with further increase in the m$_A$ values.
We notice a similar behaviour for NLO+NLL, NNLO+NNLL, NLO+$\overline{\text{NLL}}$ and NNLO+$\overline{\text{NNLL}}$.
%
%
\begin{table}[!htb]
\centering
\begin{tabularx}{1.05\textwidth}{ 
  | >{\raggedright\arraybackslash}X
  | >{\centering\arraybackslash}X
  | >{\centering\arraybackslash}X
  | >{\centering\arraybackslash}X
  | >{\centering\arraybackslash}X
  | >{\centering\arraybackslash}X
  | >{\centering\arraybackslash}X
  | >{\centering\arraybackslash}X | }
 \hline
 \textbf{Order} & \multicolumn{6}{c}{\textbf{7-point scale uncertainty}} 
 & \\ \cline{2-8} 
 & \textbf{m$_A=125$ GeV}                 
 & \textbf{m$_A=700$ GeV} 
 & \textbf{m$_A=1000$ GeV} 
 & \textbf{m$_A=1500$ GeV}   
 & \textbf{m$_A=2000$ GeV} 
 & \textbf{m$_A=2500$ GeV} 
 & \textbf{m$_A=3500$ GeV} 
 \\ \hline 
\hline
\rule{0pt}{10pt} LO        & 44.46 & 54.92 & 57.91 & 61.67 & 64.84 & 67.75 & 72.98 \\ \hline
\rule{0pt}{10pt} NLO       & 36.02 & 34.61 & 27.59 & 28.70 & 29.80 & 30.98 & 33.62 \\ \hline
\rule{0pt}{10pt} NNLO      & 21.33 & 18.18 & 16.53 & 16.23 & 16.24 & 16.41 & 16.96 \\ \hline 
\hline   
\rule{0pt}{10pt} LO+LL     & 54.49 & 61.84 & 65.13 & 69.45 & 73.29 & 76.94 & 84.43 \\ \hline
\rule{0pt}{10pt} NLO+NLL   & 45.39 & 28.45 & 32.22 & 31.15 & 30.97 & 31.23 & 32.37 \\ \hline
\rule{0pt}{10pt} NNLO+NNLL & 22.94 & 13.89 & 20.75 & 21.09 & 21.82 & 22.67 & 24.30 \\ \hline
\hline
\rule{0pt}{10pt} LO+$\overline{\text{LL}}$     
& 59.38 & 64.40 & 67.35 & 71.37 & 75.04 & 78.60 & 85.72 \\ \hline
\rule{0pt}{10pt} NLO+$\overline{\text{NLL}}$   
& 62.71 & 38.52 & 40.42 & 38.08 & 36.92 & 36.27 & 35.67 \\ \hline
\rule{0pt}{10pt} NNLO+$\overline{\text{NNLL}}$ 
& 42.11 & 26.09 & 29.31 & 28.78 & 28.94 & 29.32 & 30.02 \\ \hline
\end{tabularx}
\caption{\mk{Percentage of uncertainty in the pseudoscalar production cross-sections due to $7$-point scale variations for different values of pseudoscalar mass at 13 TeV LHC.}}
\label{tab:7pt_Comp_table}
\end{table}
}

\begin{figure}[!htb]
\centering
  \includegraphics[width=.8\linewidth]{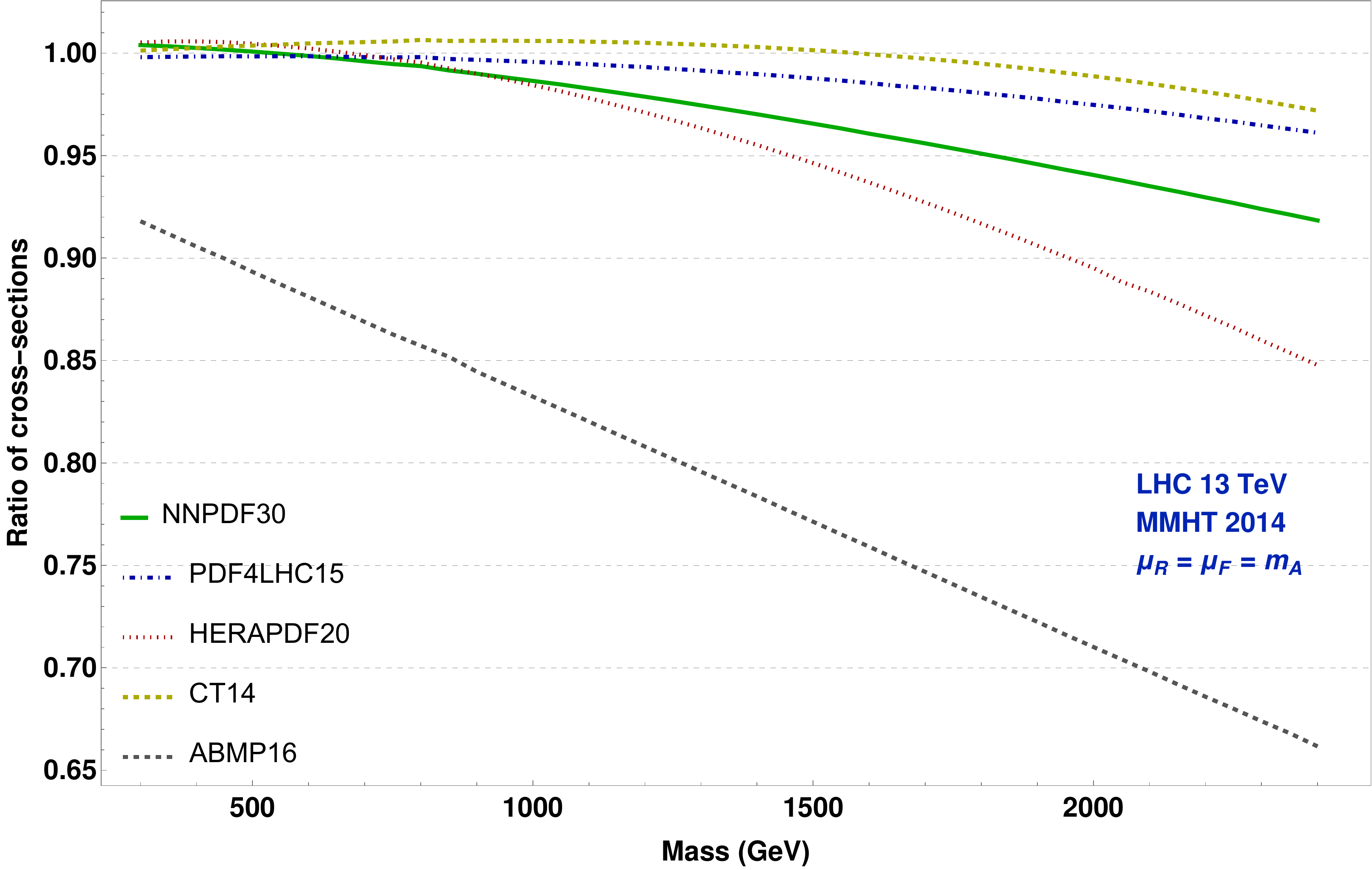}
  \caption{Pseudoscalar production cross-sections at NNLO$+\overline{\text{NNLL}}$ for different choices of PDF's, normalized w.r.t. those obtained from MMHT2014 PDF's.}
\label{fig:pdfuncertainty}
\end{figure}

For completeness, we also estimate the uncertainty due to the choice of parton densities in our calculation. 
Fig.~\ref{fig:pdfuncertainty} depicts the PDF uncertainties involved in the calculations due to the choice of different PDF sets.  
For this analysis, we choose NNPDF30, PDF4LHC15, HERAPDF20, CT14 and ABMP16 as the reference PDF's, and present the results for the production cross-section at NNLO+$\overline{\text{NNLL}}$, normalized with the corresponding results obtained from our default choice of MMHT2014 PDF set. 

\mk{
The requirement of these high-precision computations is not only significant in the discovery of the pseudoscalar Higgs but also crucial in establishing the CP properties of the discovered Higgs boson. 
As mentioned in Section \ref{sec:Introduction}, the observed Higgs boson at the LHC can be an admixture of scalar-pseudoscalar states and this mixing can be parameterized by a single mixing angle $\alpha$. In terms of this mixing angle, a pure scalar state can be represented by $\alpha=0$ and a pure pseudoscalar state by $\alpha=\pi/2$. 
The production of a mixed scalar-pseudoscalar Higgs boson from gluon fusion at the LHC has been studied in \cite{Jaquier:2019bfs} through NNLO. 
In Table \ref{tab:Mixing_table}, we give the production cross-sections at the LHC for different values of this mixing angle $\alpha$ in the $125$ GeV mass range. 
\begin{table}[!htb]
\centering
\begin{tabularx}{0.9\textwidth}{ 
  | >{\raggedright\arraybackslash}X
  | >{\centering\arraybackslash}X
  | >{\centering\arraybackslash}X
  | >{\centering\arraybackslash}X
  | >{\centering\arraybackslash}X
  | >{\centering\arraybackslash}X
  | >{\centering\arraybackslash}X
  | >{\centering\arraybackslash}X
  | >{\centering\arraybackslash}X
  | >{\raggedleft\arraybackslash}X | }
 \hline
 \vspace{0.06pt} \textbf{K-Factor} 
 & \vspace{0.06pt}$\alpha=0$ 
 & \vspace{0.06pt}$\alpha=1$ 
 & \vspace{0.06pt}$\alpha=\pi/4$ 
 & \vspace{0.06pt}$\alpha=\pi/6$
                               \\ \hline 
                               \hline
\rule{0pt}{10pt} K$_{(1)}$            & 1.6990 & 1.7124 & 1.7083 & 1.7048 \\ \hline
\rule{0pt}{10pt} K$_{(2)}$            & 2.1571 & 2.1814 & 2.1741 & 2.1677 \\ \hline
                                      \hline
\rule{0pt}{10pt} K$_{(1)}^{resum}$    & 2.0033 & 2.0803 & 2.0570 & 2.0368 \\ \hline
\rule{0pt}{10pt} K$_{(2)}^{resum}$    & 2.2785 & 2.4392 & 2.3907 & 2.3485 \\ \hline
                                      \hline 
\rule{0pt}{10pt} $\overline{K}_{(1)}^{resum}$  & 2.3425 & 2.4284 & 2.4025 & 2.3799 \\ \hline
\rule{0pt}{10pt} $\overline{K}_{(2)}^{resum}$  & 2.4737 & 2.5966 & 2.5595 & 2.5272 \\ \hline
\end{tabularx}
\caption{\mk{K-factors for the production cross-sections of a mixed state of scalar and pseudoscalar Higgs bosons. 
The production cross-sections are computed for $m_A=125$ GeV at 13 TeV LHC using MMHT2014 PDFs for the central scale choice 
of $\mu_R=\mu_F=m_A$.  Each column represents different values of the mixing angle $\alpha$.
}}
\label{tab:Mixing_table}
\end{table}
%
Here, we observe that the inclusion of the SV resummed results increase the NLO (NNLO) cross-sections by about $30\%~(12\%)$ of the LO ones while a further addition of the NSV resummed results, increase the NLO+NLL (NNLO+NNLL) cross-sections by about $33\%~(20\%)$ of LO ones for the pure scalar state.
However for the pure pseudoscalar state, we observe that the inclusion of the SV resummed results increase the NLO (NNLO) cross-sections by about $37\%~(26\%)$ of the LO ones while a further addition of the NSV resummed results, increase the NLO+NLL (NNLO+NNLL) cross-section by approximately $35\%~(16\%)$ of the LO ones.
For the mixed cases considered, the inclusion of the SV resummed results increase the NLO (NNLO) cross-sections by about $35\%~(22\%)$ of the LO ones for $\alpha=\pi/4$ and by approximately $33\%~(18\%)$ of the LO ones for $\alpha=\pi/6$. 
Similarly, the addition of the NSV resummed results increase the NLO+NLL (NNLO+NNLL) cross-sections by approximately $35\%~(17\%)$ of the LO ones for $\alpha=\pi/4$ and by about $34\%~(18\%)$ of the LO ones for $\alpha=\pi/6$. 
Overall, we observe that as we change the mixing angle, the corresponding QCD corrections change only by a few percent. 

In \cite{Jaquier:2019bfs}, the authors also conclude that if a Higgs boson production is considered for any arbitrary value of the mixing angle while neglecting its decay, then the results up to NNLO may be obtained by a simple rescaling of the scalar and pseudoscalar cross-sections as below:
\begin{equation}
 \sigma = \cos^2\alpha \cdot \sigma_H + \sin^2\alpha \cdot \sigma_A
 \label{eqn:mixing}
\end{equation} 
where $\sigma$ is the superposed cross-section, $\sigma_H$ is the pure Higgs cross-section and $\sigma_A$ is the pure pseudoscalar Higgs cross-section. 
In such a scenario, if the pseudoscalar Higgs boson production cross-section is made available to a precision comparable to that of the scalar Higgs boson, then it can prove helpful in extracting the mixing angle to a better accuracy.

It is necessary to clarify here that a number of angular observables corresponding to the decay products of the Higgs boson must be studied for establishing its properties. When such a Higgs decay is considered, the simple reweighting formula in eqn. \ref{eqn:mixing} fails, and consequently, the corresponding K-factors similar to those given in Table \ref{tab:Mixing_table} get modified slightly. However, such a detailed analysis (including the angular distributions of the decay products) is beyond the scope of this article.
}

\begin{figure}[!htb]
\centering
\begin{subfigure}{.5\textwidth}
  \centering
  \includegraphics[width=1.0\linewidth]{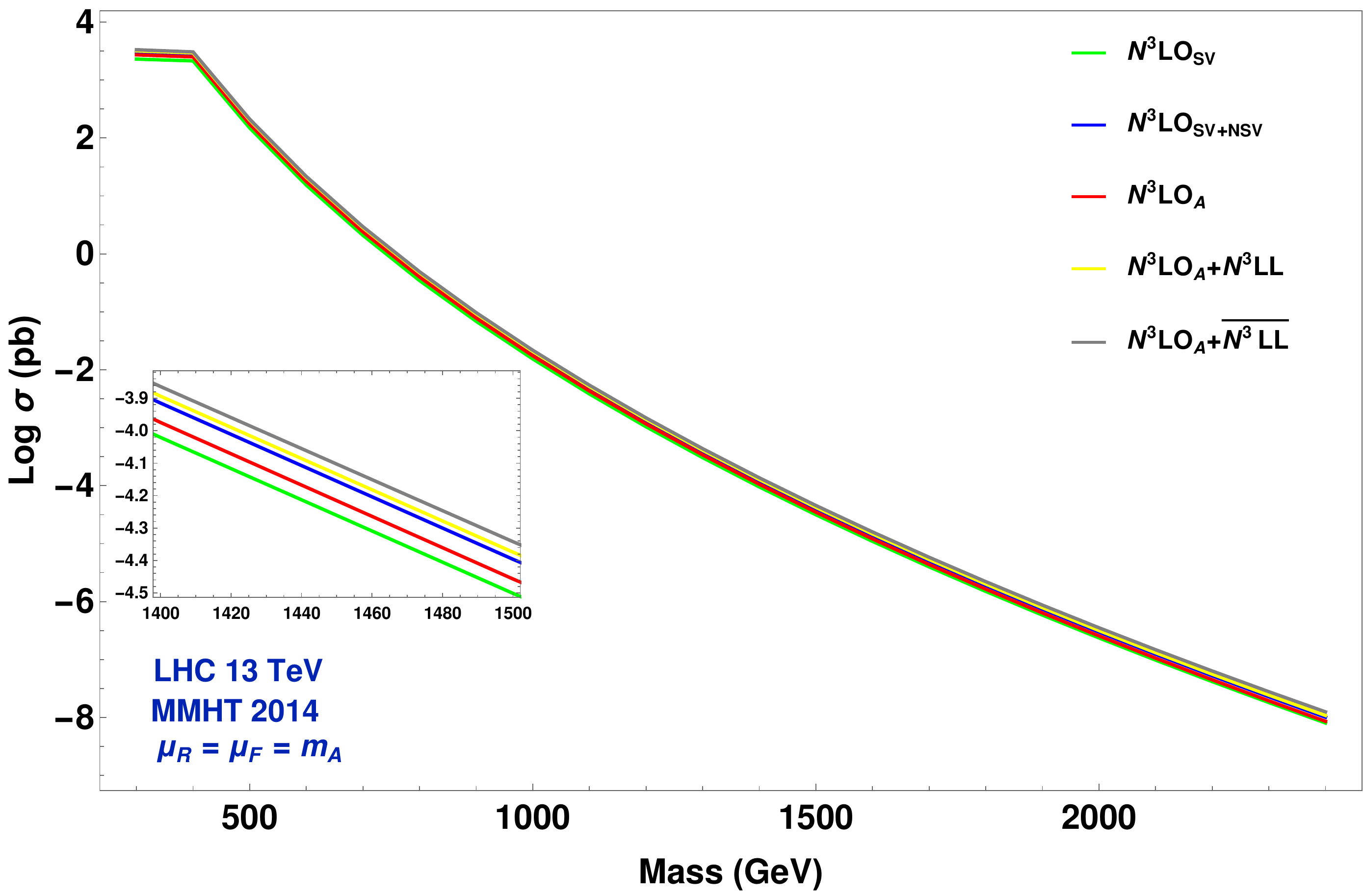}
  \label{fig:sub3}
\end{subfigure}%
\begin{subfigure}{.5\textwidth}
  \centering
  \includegraphics[width=1.0\linewidth]{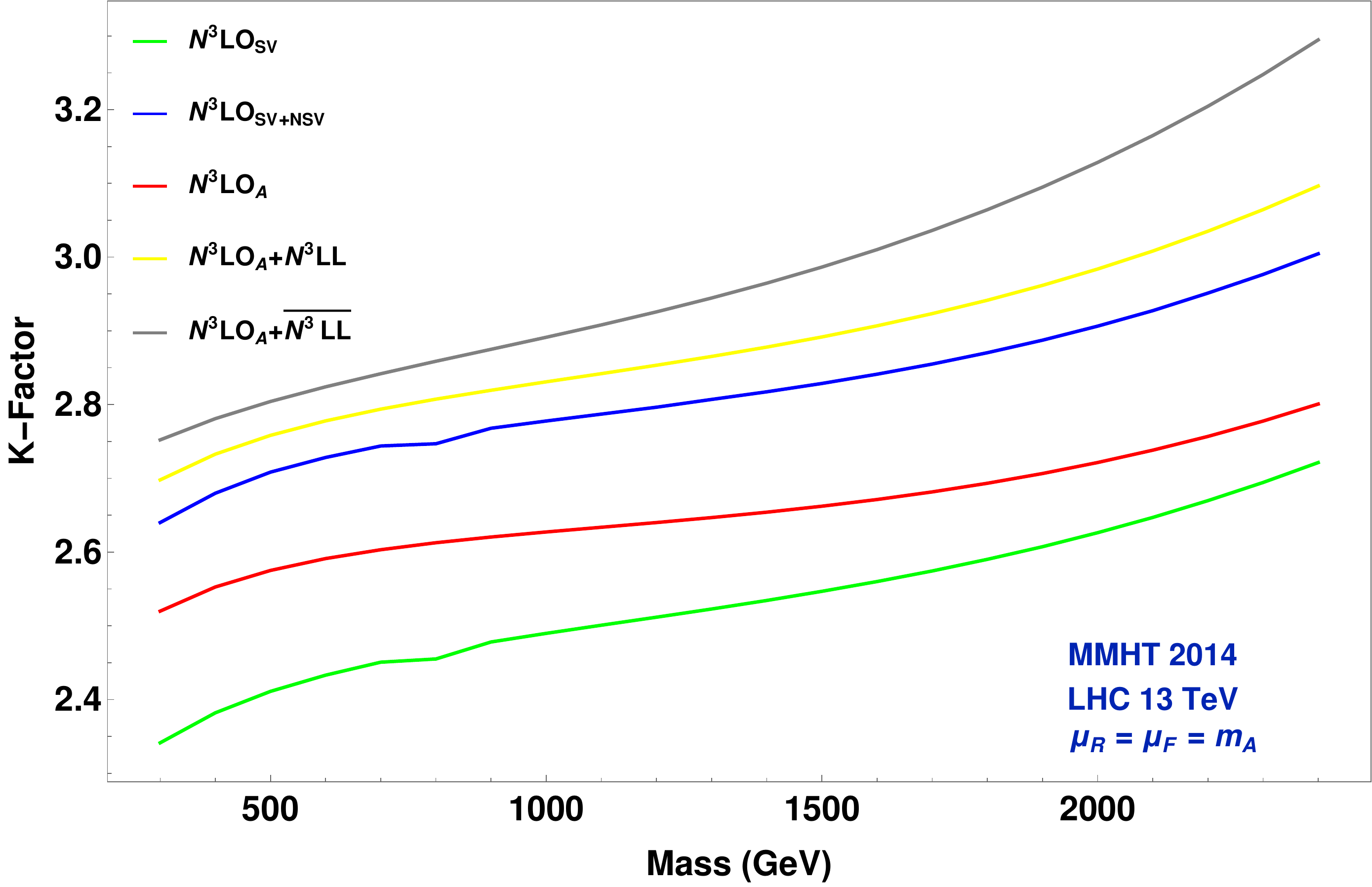}
  \label{fig:sub4}
\end{subfigure}
\caption{Pseudoscalar production cross-section at third order in QCD (left panel) and the corresponding K-Factors (right panel). The corresponding resummed results to N$^3$LL accuracy are also given after matching them to the approximate full N$^3$LO$_A$ results.}
\label{fig:crossK3}
\end{figure}

Finally, we attempt to give predictions for the pseudoscalar production cross-section after including the resummed results to $\overline{\text{N$^3$LL}}$ accuracy and then matching them to the FO N$^3$LO$_A$ results. 
Here, N$^3$LO$_A$ results represent the approximate full FO results at third order in QCD which we have taken from the public code 
\cite{Ball_2013,Bonvini_2014_TROLL,Bonvini_2016_TROLL,Bonvini_2018,Bonvini_2018_1}.
In the left panel of Fig.~\ref{fig:crossK3}, we plot the pseudoscalar production cross-section as a function of its mass m$_A$ at third order and present the results upto N$^3$LO$_A+\overline{\text{N$^3$LL}}$. 
In the right panel, we give the corresponding K-factors obtained by normalizing with the LO cross-sections.
\mk{We also give K-factors for the FO case by keeping only the third order SV and SV+NSV results, and compare with those of the N$^3$LO$_A$. 
The N$^3$LO$_\text{SV}$ results are closer to the N$^3$LO$_A$ ones in the high mass region, while N$^3$LO$_\text{SV+NSV}$ results get a bit closer to the N$^3$LO$_A$ ones in the low mass region. 
We observe that the inclusion of NSV corrections either at the FO level or their resummation through NSV substantially increase the cross-sections. 
}

However, a detailed phenomenological study at $\overline{\text{N$^3$LL}}$ level accuracy, including the estimation of theoretical uncertainties, require additional terms such as the resummation constant $h_{33}^g(\omega)$ that are yet to be determined.
In each of the above cases, the N$^3$LO level PDF's are also not available yet. 
However, it can be anticipated that the uncertainty will get further reduced if the full N$^3$LO results with the corresponding PDF's are made available.
%
%
%

\section{Summary}
\label{sec:Summary}

In this work, we have performed the NSV resummation for the pseudoscalar production process to $\overline{\text{NNLL}}$ accuracy in QCD at the LHC. 
We further make a detailed phenomenology of the same and present our results for 13 TeV LHC.
We have computed the NSV corrections at both first and second orders in QCD and compared them with the full NLO and NNLO corrections for gluon fusion sub-process. 
We find that these NSV corrections are potentially large and enhance the pseudoscalar production cross-sections much more than the conventional SV or threshold logarithms.
We also give numerical results for the NSV resummed results to $\overline{\text{NNLL}}$ accuracy by systematically matching to the FO NNLO cross-sections. 
We find that these NSV resummed predictions give large contributions to the cross-sections than what the SV resummation results do. 
We estimate the size of these corrections in terms of the resummed K-factors defined 
{\it{w.r.t.}} the LO and find them to be as large as $3$ at NNLO+$\overline{\text{NNLL}}$ accuracy in the high mass region. 
We also estimate in our predictions the uncertainties due to the choice of various parton densities and those due to the unknown renormalization and factorization scales. 
We find that the conventional 7-point scale uncertainties do not get improvement
after performing the NSV resummation, suggesting the requirement of including NSV contributions from other parton channels as well as beyond NSV contributions in the gluon fusion channel.
Specifically, we notice that the $\mu_F$ scale variations while keeping $\mu_R$ fixed, lead to large uncertainties in the pseudoscalar production cross-sections in the high mass region.
\mk{
However, for the pure $\mu_R$ variation, keeping $\mu_F$ fixed, we find that the scale uncertainties get reduced significantly to about $1.8\%$ for $m_A=700$ GeV and are much smaller than those of the SV resummed results $(\sim 6.4\%)$.
We also present the production cross-sections for mixed scalar-pseudoscalar state and study the impact of QCD corrections on them for different values of the mixing angle $\alpha$. We find that these QCD corrections change with the mixing angle by a few percent. We anticipate that these precision results are useful in future analyses aiming to study the CP-properties of the discovered Higgs boson.
}

\newpage
\section*{Acknowledgments}
\label{sec:Acknowledgements}

We would like to thank Surabhi Tiwari for fruitful discussions and her extensive support in providing various other relevant data for comparison. 
We would also like to thank Aparna Sankar for highlighting significant points and relevant discussions.
AB would like to thank Pooja Mukherjee for useful discussions. 
AB is funded by the Department of Atomic Energy (DAE), India.

\clearpage
\newpage
\appendix
\section*{Appendix}
\addcontentsline{toc}{section}{Appendices}
\section{NSV coefficient functions}
\label{appendix:A}

In this section, we present our analytical results of the SV+NSV coefficient functions up to N$^3$LO order. 
Expanding the SV+NSV coefficient function in eqn.~\ref{eq:deltaA}, in powers of $a_s$, we obtain
\begin{equation}
\Delta_{g}^{A,NSV}\left(z,q^{2}\right)=\sum_{i=0}^{\infty}a_{s}^{i}~\Delta_{g,i}^{A,NSV}\left(z,q^{2}\right).
\end{equation}
where
\begin{align*}
 \Delta_{g,0}^{A,NSV} &= \delta(1-z),
                            \nonumber\\
 \Delta_{g,1}^{A,NSV} &= \delta(1-z) \bigg[8 C_A \big(1 + \zeta_2 \big) \bigg] +
                            \mathcal{D}_1 \bigg[16 C_A \bigg] - \bigg[16 C_A\bigg] \log(1-z) + 8 C_A,
                            \nonumber\\
 \Delta_{g,2}^{A,NSV} &= \delta(1-z) \bigg[C_A^2\bigg( + \dfrac{494}{3} -
                         \dfrac{220\zeta_3}{3} - \dfrac{4\zeta_2^2}{5} + \dfrac{1112\zeta_2}{9}\bigg) 
                         \nonumber\\
                         &+ C_A n_f \bigg( - \dfrac{82}{3} - \dfrac{80\zeta_2}{9} - \dfrac{8\zeta_3}{3} \bigg) + C_F n_f \bigg( - \dfrac{160}{3} + 12 \log\dfrac{\mu_R^2}{m_t^2} + 16 \zeta_3\bigg)\bigg]
                         \nonumber\\
                         & + \mathcal{D}_0 \bigg[ C_A^2 \bigg( - \dfrac{1616}{27} 
                         + \dfrac{176 \zeta_2}{3} + 312 \zeta_3 \bigg)
                         + C_A n_f \bigg(\dfrac{224}{27}  - 
                         \dfrac{32 \zeta_2}{3} \bigg) \bigg]
                         \nonumber\\
                         &+ \mathcal{D}_1 \bigg[ C_A^2 \bigg( + \dfrac{2224}{9} 
                         - 160 \zeta_2\bigg) - C_A n_f \bigg(\dfrac{160}{9}\bigg) \bigg]
                        \nonumber\\
                         & + \mathcal{D}_2 \bigg[-C_A^2 \bigg(\dfrac{176}{3}\bigg) +C_A n_f\bigg(\dfrac{32}{3}\bigg)\bigg] + \mathcal{D}_3 \bigg[C_A^2\bigg(128\bigg)\bigg] - \bigg[C_A^2\bigg(128\bigg)\bigg] \log^3(1-z)
                         \nonumber\\
                         & + \bigg[C_A^2\bigg(\dfrac{920}{3}\bigg) - C_A n_f \bigg(\dfrac{32}{3}\bigg)\bigg] \log^2(1-z) + \bigg[C_A^2\bigg(-\dfrac{2740}{9} + 160 \zeta_2\bigg) \nonumber\\
                         & + C_A n_f \bigg(\dfrac{244}{9} \bigg)\bigg] \log(1-z)
                          + \bigg[C_A^2 \bigg(\dfrac{4340}{27} - \dfrac{608 \zeta_2}{3} - 312\zeta_3 \bigg) - C_A n_f \bigg( \dfrac{392}{27}  + \dfrac{32 \zeta_2}{3} \bigg) \bigg],
                         \nonumber\\
 \Delta_{g,3}^{A,NSV} &= 512~ C_A^3 \mathcal{D}_5 - 512~ C_A^3 \log^5(1-z) + 
                         \log^4(1-z) \bigg[\dfrac{22592~ C_A^3}{9} - \dfrac{1280~ C_A^2 n_f}{9}\bigg] 
                         \nonumber \\
                         & + \mathcal{D}_4 \bigg[-\dfrac{7040~ C_A^3}{9} + \dfrac{1280~ C_A^2 n_f}{9} \bigg] + \mathcal{D}_3 \bigg[-\dfrac{10496~ C_A^2 n_f}{27} + \dfrac{256~ C_A n_f^2}{27} + C_A^3 \bigg(\dfrac{86848}{27} - 3584~ \zeta_2\bigg)\bigg]  \nonumber \\
                         & + \log^3(1-z) \bigg[\dfrac{6560~ C_A^2 n_f}{9} - \dfrac{256~ C_A n_f^2}{27} + C_A^3 \bigg(-\dfrac{138656}{27} + 3584~ \zeta_2\bigg) \bigg] 
                         + C_A \bigg[ n_f^2 \bigg(\dfrac{3136}{29}  
                         - \dfrac{304~ \zeta_2}{9} - \dfrac{320~ \zeta_3}{27} \bigg)
                         \nonumber \\
                         & + C_F n_f \bigg(-\dfrac{16826}{27} 
                         + 96~ \log\dfrac{\mu_R^2}{m_t^2} + \dfrac{80~ \zeta_2}{3} 
                         + \dfrac{64~ \zeta_2^2}{5} + \dfrac{2384~ \zeta_3}{9} \bigg) \bigg]
                         + C_A^2 n_f \bigg(-\dfrac{479636}{729} + \dfrac{79096~ \zeta_2}{81} + \dfrac{272~ \zeta_2^2}{5} 
                         \nonumber \\
                         & + \dfrac{13120~ \zeta_3}{9} \bigg)
                         + \log^2(1-z) \bigg[C_A \bigg(-32~ C_F n_f 
                         + \dfrac{944~ n_f^2}{27} \bigg) + C_A^2 n_f \bigg( - \dfrac{34984}{27} + \dfrac{2180~ \zeta_2}{3} \bigg) 
                         \nonumber \\
                         & + C_A^3 \bigg( \dfrac{239744}{27} - \dfrac{34352~ \zeta_2}{3} - 11584 ~\zeta_3 \bigg)\bigg]  
                         + \mathcal{D}_2 \bigg[C_A \bigg(32~ C_F n_f - \dfrac{640~ n_f^2}{27} + C_A^2 n_f \bigg(\dfrac{16928}{27} - \dfrac{2176 ~\zeta_2}{3} \bigg)
                         \nonumber \\
                         & + C_A^3 \bigg(-\dfrac{79936}{27} + \dfrac{11968~ \zeta_2}{3} + 11584~ \zeta_3 \bigg) \bigg] 
                        + \log(1-z) \bigg[ C_A \bigg\{ n_f^2 \bigg(-\dfrac{2608}{81} + \dfrac{256~ \zeta_2}{9}\bigg) 
                        \nonumber \\
                         & + C_F n_f \bigg(1040 - 192 \log\dfrac{\mu_R^2}{m_t^2} - \dfrac{16~ \zeta_2}{3} - 384~ \zeta_3 \bigg)\bigg\}  
                         + C_A^2 n_f \bigg( \dfrac{125536}{81} - \dfrac{14480~ \zeta_2}{9} - \dfrac{2992~ \zeta_3}{3} \bigg) 
                         \nonumber \\                         
                         \end{align*}
                         \begin{align}
                         & + C_A^3 \bigg( - \dfrac{221824}{27} + \dfrac{87520~ \zeta_2}{9} + \dfrac{9856~ \zeta_2^2}{5} + 23168~ \zeta_3 \bigg) \bigg] 
                         + \mathcal{D}_1 \bigg[C_A^3 \bigg( \dfrac{414616}{81} - \dfrac{13568~ \zeta_2}{3} - \dfrac{9856~ \zeta_2^2}{5} 
                         \nonumber \\
                         & - \dfrac{22528~ \zeta_3}{3}) + C_A^2 n_f \bigg(-\dfrac{79760}{81} + \dfrac{6016~ \zeta_2}{9} + \dfrac{2944~ \zeta_3}{3} \bigg) + C_A \bigg\{ n_f^2 \bigg( \dfrac{1600}{81} - \dfrac{256~ \zeta_2}{9}\bigg) 
                         \nonumber \\
                         & + C_F n_f \bigg(-1000 + 192~ \log\dfrac{\mu_R^2}{m_t^2} + 384~ \zeta_3 \bigg) \bigg\} \bigg] 
                         + C_A^3 \bigg(2650990/729 - \dfrac{489184~ \zeta_2}{81} - \dfrac{26272~ \zeta_2^2}{15} - \dfrac{330184~ \zeta_3}{27} 
                         \nonumber \\
                         & + \dfrac{23200~ \zeta_2~ \zeta_3}{3} - 11904~ \zeta_5 \bigg) 
                         + \mathcal{D}_0 \bigg[ C_A^2 n_f \bigg( \dfrac{173636}{729} - \dfrac{41680~ \zeta_2}{81} - \dfrac{544~ \zeta_2^2}{15} - \dfrac{7600~ \zeta_3}{9} \bigg) 
                         + C_A \bigg \{ C_F~ n_f \bigg(\dfrac{3422}{27} 
                         \nonumber\\
                         & - 32 ~\zeta_2 - \dfrac{64~ \zeta_2^2}{5} - \dfrac{608~ \zeta_3}{9} \bigg) 
                         + n_f^2 \bigg( -\dfrac{3712}{729} + \dfrac{640~ \zeta_2}{27} + \dfrac{320~ \zeta_3}{27} \bigg) \bigg\} 
                         + C_A^3 \bigg( - \dfrac{943114}{729} + \dfrac{175024~ \zeta_2}{81}
                         \nonumber\\                    
                         & + \dfrac{4048~ \zeta_2^2}{15} + \dfrac{210448 \zeta_3}{27} - \dfrac{23200~ \zeta_2~ \zeta_3}{3} + 11904~ \zeta_5 \bigg) \bigg]
                         \nonumber \\
                         & +  \delta(1-z) \bigg[-4 n_f C_J^{(2)}~ \log^2(1-z) 
                         + C_F n_f^2 \bigg( \dfrac{1498}{9} - \dfrac{40~ \zeta_2}{9} - \dfrac{32~ \zeta_2^2}{45} - \dfrac{224~ \zeta_3}{3} \bigg) 
                         \nonumber \\
                         & + C_A^3 \bigg( \dfrac{114568}{27} + \dfrac{266155~ \zeta_2}{162} - \dfrac{4007~ \zeta_2^2}{10} - \dfrac{64096~ \zeta_2^3}{105} - 3932~ \zeta_3 + \dfrac{7832~ \zeta_2 ~\zeta_3}{3} + \dfrac{13216~ \zeta_3^2}{3} - \dfrac{30316~ \zeta_5}{9} \bigg) 
                         \nonumber \\
                         & + C_F^2~ n_f \bigg( \dfrac{457}{3} + 208~ \zeta_3 - 320~ \zeta_5 \bigg) + C_A^2~ n_f \bigg( -\dfrac{113366}{81} 
                         - \dfrac{56453~ \zeta_2}{405} + \dfrac{21703~ \zeta_2^2}{135} + \dfrac{8840~ \zeta_3}{27} - \dfrac{2000~ \zeta_2~ \zeta_3}{3} + \dfrac{6952~ \zeta_5}{9} \bigg) 
                         \nonumber \\
                         & + C_A \bigg \{ n_f^2 \bigg( \dfrac{6914}{81} - \dfrac{7088~ \zeta_2}{405} - \dfrac{2288~ \zeta_2^2}{135} 
                         + \dfrac{688~ \zeta_3}{27} \bigg) 
                         + C_F n_f \bigg(-1797 + 96~ \log\dfrac{\mu_R^2}{m_t^2} - \dfrac{4160~ \zeta_2}{9} + 96~ \log\dfrac{\mu_R^2}{m_t^2} ~\zeta_2 
                         \nonumber \\
                         & + \dfrac{176~ \zeta_2^2}{45} + \dfrac{1856~ \zeta_3}{3} + 192~ \zeta_2~ \zeta_3 + \dfrac{3872~ \zeta_5}{9} \bigg) \bigg\} \bigg].
\end{align}

\section{SV coefficients in soft collinear distribution}
\label{appendix:B}

In this section, we present the explicit expressions for $\hat{\phi}_{g}^{SV,\left(i\right)}\left(\varepsilon\right)$ appearing in Eqn. (\ref{eq:phicapg}):
\begin{align}
 \hat{\phi}_{g}^{SV,\left(1\right)}\left(\varepsilon\right) &= 
 -\dfrac{3}{16} C_A \varepsilon^2 \zeta_2^2+\frac{8
    C_A}{\varepsilon^2}+\frac{7 C_A \zeta_3}{3}\varepsilon-3
    C_A \zeta_2,
    \nonumber \\
  \hat{\phi}_{g}^{SV,\left(2\right)}\left(\varepsilon\right) &=   
    \varepsilon \bigg\{\frac{3}{16} \beta_0 C_A \zeta_2^2+\frac{11
    C_A^2 \zeta_2^2}{80}-\frac{203}{6} C_A^2 \zeta_2
    \zeta_3+\frac{707 C_A^2 \zeta_2}{27}+\frac{2077 C_A^2
    \zeta_3}{54}+\frac{43 C_A^2 \zeta_5}{2}
    \nonumber \\
   & -\frac{3644
    C_A^2}{243}-\frac{1}{40} C_A n_f \zeta_2^2-\frac{98
    C_A n_f \zeta_2}{27}-\frac{155 C_A n_f
    \zeta_3}{27}+\frac{488 C_A n_f}{243}\bigg\}
    +\bigg\{ 2
    C_A^2 \zeta_2^2
    \nonumber \\
    &-\frac{469 C_A^2 \zeta_2}{18}-\frac{88
    C_A^2 \zeta_3}{3}+\frac{1214 C_A^2}{81}
    +\frac{35 C_A
    n_f \zeta_2}{9}+\frac{16 C_A n_f
    \zeta_3}{3}-\frac{164 C_A n_f}{81}-\frac{7 \beta_0 C_A
    \zeta_3}{3}\bigg\}
    \nonumber \\
    &  +\dfrac{1}{\varepsilon}\bigg\{6
    \beta_0 C_A \zeta_2
    +\frac{11 C_A^2 \zeta_2}{3}+14
    C_A^2 \zeta_3-\frac{404 C_A^2}{27}-\frac{2 C_A
    n_f \zeta_2}{3}+\frac{56 C_A
    n_f}{27}\bigg\}
    \nonumber \\
    &+\frac{1}{\varepsilon^2}\bigg\{-4 C_A^2 \zeta_2+\frac{134
    C_A^2}{9}-\frac{20 C_A n_f}{9}\bigg\}-\frac{4 \beta_0
    C_A}{\varepsilon^3},
    \nonumber \\
\hat{\phi}_{g}^{SV,\left(3\right)}\left(\varepsilon\right) &=
    \frac{32
    \beta_0^2 C_A}{9 \varepsilon^4}
    +\dfrac{1}{\varepsilon^3}\bigg\{\frac{64}{9} \beta_0 C_A^2
    \zeta_2-\frac{2144 \beta_0 C_A^2}{81}+\frac{320 \beta_0
    C_A n_f}{81}-\frac{8 \beta_1
    C_A}{9}\bigg\}
    \nonumber \\
    &\dfrac{1}{\varepsilon^2}\bigg\{-12 \beta_0^2 C_A \zeta_2-\frac{88}{9} \beta_0
    C_A^2 \zeta_2-\frac{112}{3} \beta_0 C_A^2
    \zeta_3+\frac{3232 \beta_0 C_A^2}{81}+\frac{16}{9} \beta_0
    C_A n_f \zeta_2
    \nonumber \\
    &-\frac{448 \beta_0 C_A
    n_f}{81}+\frac{352 C_A^3 \zeta_2^2}{45}-\frac{2144
    C_A^3 \zeta_2}{81}+\frac{176 C_A^3
    \zeta_3}{27}+\frac{980 C_A^3}{27}+\frac{320}{81} C_A^2
    n_f \zeta_2
    \nonumber \\
    &-\frac{224}{27} C_A^2 n_f
    \zeta_3-\frac{1672 C_A^2 n_f}{243}+\frac{64}{9} C_A
    C_f n_f \zeta_3-\frac{220 C_A C_f
    n_f}{27}-\frac{32 C_A n_f^2}{243}\bigg\}    
    \nonumber \\
    &+\dfrac{1}{\varepsilon}\bigg\{-8 \beta_0 C_A^2
    \zeta_2^2+\frac{938}{9} \beta_0 C_A^2
    \zeta_2+\frac{352}{3} \beta_0 C_A^2 \zeta_3-\frac{4856
    \beta_0 C_A^2}{81}-\frac{140}{9} \beta_0 C_A n_f
    \zeta_2
    \nonumber \\
    &-\frac{64}{3} \beta_0 C_A n_f
    \zeta_3+\frac{656 \beta_0 C_A n_f}{81}+3 \beta_1
    C_A \zeta_2-\frac{352}{15} C_A^3
    \zeta_2^2-\frac{176}{9} C_A^3 \zeta_2
    \zeta_3+\frac{12650 C_A^3 \zeta_2}{243}
    \nonumber \\
    &+\frac{1316
    C_A^3 \zeta_3}{9}-64 C_A^3 \zeta_5-\frac{136781
    C_A^3}{2187}+\frac{32}{5} C_A^2 n_f
    \zeta_2^2-\frac{2828}{243} C_A^2 n_f
    \zeta_2-\frac{728}{81} C_A^2 n_f \zeta_3
    \nonumber \\
    &+\frac{11842
    C_A^2 n_f}{2187}-\frac{32}{15} C_A C_f n_f
    \zeta_2^2-\frac{4}{3} C_A C_f n_f
    \zeta_2-\frac{304}{27} C_A C_f n_f
    \zeta_3+\frac{1711 C_A C_f n_f}{81}
    \nonumber \\
    &+\frac{40}{81}
    C_A n_f^2 \zeta_2-\frac{112}{81} C_A n_f^2
    \zeta_3+\frac{2080 C_A
    n_f^2}{2187}\bigg\}+\bigg\{-\frac{11}{30} \beta_0 C_A^2
    \zeta_2^2+\frac{812}{9} \beta_0 C_A^2 \zeta_2
    \zeta_3
    \nonumber \\
    &-\frac{5656}{81} \beta_0 C_A^2
    \zeta_2
    -\frac{8308}{81} \beta_0 C_A^2
    \zeta_3-\frac{172}{3} \beta_0 C_A^2 \zeta_5+\frac{29152
    \beta_0 C_A^2}{729}+\frac{1}{15} \beta_0 C_A n_f
    \zeta_2^2
    \nonumber \\
    &+\frac{1240}{81} \beta_0 C_A n_f
    \zeta_3-\frac{3904 \beta_0 C_A n_f}{729}+\frac{1}{16}
    \beta_1 C_A \varepsilon \zeta_2^2-\frac{7 \beta_1 C_A
    \zeta_3}{9}+\frac{152 C_A^3 \zeta_2^3}{189}+\frac{1964
    C_A^3 \zeta_2^2}{27}
    \nonumber \\
    &+\frac{11000}{27} C_A^3 \zeta_2
    \zeta_3-\frac{765127 C_A^3 \zeta_2}{1458}+\frac{536
    C_A^3 \zeta_3^2}{9}-\frac{59648 C_A^3
    \zeta_3}{81}-\frac{1430 C_A^3 \zeta_5}{9}+\frac{7135981
    C_A^3}{26244}
    \nonumber \\
    &-\frac{532}{27} C_A^2 n_f
    \zeta_2^2-\frac{1208}{27} C_A^2 n_f \zeta_2
    \zeta_3+\frac{105059}{729} C_A^2 n_f
    \zeta_2+\frac{45956}{243} C_A^2 n_f
    \zeta_3+\frac{148}{9} C_A^2 n_f \zeta_5
    \nonumber \\
    &-\frac{716509
    C_A^2 n_f}{13122}+\frac{152}{45} C_A C_f n_f
    \zeta_2^2-\frac{88}{3} C_A C_f n_f \zeta_2
    \zeta_3+\frac{605}{18} C_A C_f n_f
    \zeta_2+\frac{2536}{81} C_A C_f n_f
    \zeta_3
    \nonumber \\
    &+\frac{112}{9} C_A C_f n_f
    \zeta_5-\frac{42727 C_A C_f n_f}{972}+\frac{32}{27}
    C_A n_f^2 \zeta_2^2-\frac{1996}{243} C_A n_f^2
    \zeta_2-\frac{2720}{243} C_A n_f^2
    \zeta_3
    \nonumber \\
    &+\frac{11584 C_A n_f^2}{6561}-\frac{1}{4} \beta_0^2 C_A
    \zeta_2^2+\frac{784}{81} \beta_0 C_A n_f
    \zeta_2\bigg\}.
\end{align}

\section{Singular NSV coefficients in soft collinear distribution}
\label{appendix:C}

In this section, we present the explicit expressions for the singular coefficients $\varphi_{s,g}^{NSV,\left(i\right)}\left(z,\varepsilon\right)$ appearing in Eqn. (\ref{eq:phiNSV}):
\begin{align}
\varphi_{s,g}^{NSV,\left(1\right)}\left(z,\varepsilon\right) &= -\dfrac{8 C_A}{\varepsilon},
\nonumber \\
\varphi_{s,g}^{NSV,\left(2\right)}\left(z,\varepsilon\right) &= \dfrac{8 \beta_0 C_A}{\varepsilon^2} + \dfrac{1}{\varepsilon}\textbf{\bigg\{}
    C_A^2 \left(8 \zeta_2-\frac{268}{9}\right)+\frac{40
    C_A n_f}{9}+16 C_A^2~\text{log}{\text{$(1-z)$}}\bigg\},
    \nonumber \\
\varphi_{s,g}^{NSV,\left(3\right)}\left(z,\varepsilon\right) &=  
-\dfrac{32 \beta_0^2 C_A }{3 \varepsilon^3}+
\dfrac{1}{\varepsilon^2}\bigg\{\frac{8
    \beta_1 C_A }{3}-\frac{8}{3} \beta_0 \bigg(
    C_A^2 \left(8 \zeta_2-\frac{268}{9}\right)+\frac{40
    C_A n_f}{9}
    \nonumber \\
    & +16 C_A^2~
    \text{log}{\text{$(1-z)$}}\bigg)\bigg\}
    + \dfrac{2}{3 \varepsilon}\bigg\{ C_A^3
    \left(-\frac{176 \zeta_2^2}{5}+\frac{1072 \zeta_2}{9}+\frac{56
    \zeta_3}{3}-166\right)
    \nonumber \\
    & +\frac{80}{3} C_A^2 n_f T_f
     +C_A^2 n_f \left(-\frac{160 \zeta_2}{9}+\frac{112
    \zeta_3}{3}+\frac{548}{27}\right)+16 C_A C_f n_f
    T_f
    \nonumber \\
    & +C_A C_f n_f \left(\frac{86}{3}-32
    \zeta_3\right) +\frac{16 C_A n_f^2}{27}+\text{log}{\text{$(1-z)$}}
    \bigg(C_A^3 \left(\frac{2144}{9}-64 \zeta_2\right)
    \nonumber \\
    & -\frac{320
    C_A^2 n_f}{9}\bigg)\bigg\}.
\end{align}



\clearpage
\bibliography{ggA_NSV}
\end{document}